\documentclass[sigconf]{acmart}

\copyrightyear{2026}
\acmYear{2026}
\setcopyright{cc}
\setcctype{by}
\acmConference[WWW '26]{Proceedings of the ACM Web Conference 2026}{April 13--17, 2026}{Dubai, United Arab Emirates}
\acmBooktitle{Proceedings of the ACM Web Conference 2026 (WWW '26), April 13--17, 2026, Dubai, United Arab Emirates}
\acmPrice{}
\acmDOI{10.1145/3774904.3792458}
\acmISBN{979-8-4007-2307-0/2026/04}

\usepackage{amsmath,amsfonts}
\usepackage{algorithmic}
\usepackage{graphicx}
\usepackage{textcomp}
\usepackage{enumitem}
\usepackage{xcolor}
\usepackage{framed}
\usepackage{xspace}
\usepackage{booktabs}
\usepackage[most]{tcolorbox}
\usepackage{subfigure}
\usepackage{multirow}
\usepackage{algorithm}
\usepackage{colortbl}
\usepackage{url}
\usepackage{booktabs}
\usepackage{fvextra}
\usepackage{diagbox}
\newlength{\thinline}
\newlength{\thickline}
\definecolor{mygray}{RGB}{199, 216, 230}
\definecolor{mydeepgray}{RGB}{140, 155, 190}
\definecolor{mytintgray}{RGB}{227, 235, 242}
\definecolor{darkgreen}{rgb}{0.0, 0.5, 0.0}

\newcommand{\name}{\textsc{ARuleCon}\xspace}
\usepackage{tabularx}
\usepackage{varwidth}
\DefineVerbatimEnvironment{ArxivMinted}{Verbatim}{
  breaklines,
  fontsize=\small,
  frame=single
}
\NewDocumentEnvironment{minted}{O{}m}
  {\VerbatimEnvironment\begin{ArxivMinted}}
  {\end{ArxivMinted}}
\def\BibTeX{{\rm B\kern-.05em{\sc i\kern-.025em b}\kern-.08em
    T\kern-.1667em\lower.7ex\hbox{E}\kern-.125emX}}

\begin{document}

\title{\name: Agentic Security Rule Conversion}

\author{Ming Xu\textsuperscript{\#}}
\affiliation{%
  \institution{National University of Singapore}
  \city{Singapore}
  \country{Singapore}
}
\email{mingxu@nus.edu.sg}

\author{Hongtai Wang}
\affiliation{%
  \institution{National University of Singapore}
  \city{Singapore}
  \country{Singapore}
}
\email{e1132287@nus.edu.sg}

\author{Yanpei Guo}
\affiliation{%
  \institution{National University of Singapore}
  \city{Singapore}
  \country{Singapore}
}
\email{guo.yanpei@nus.edu.sg}

\author{Zhengmin Yu}
\affiliation{%
  \institution{Fudan University}
  \city{Shanghai}
  \country{China}
}
\email{zmyu23@m.fudan.edu.cn}

\author{Weili Han}
\affiliation{%
  \institution{Fudan University}
  \city{Shanghai}
  \country{China}
}
\email{wlhan@fudan.edu.cn}

\author{Hoon Wei Lim}
\affiliation{%
  \institution{Cyber Special Ops-R\&D, NCS Group}
  \city{Singapore}
  \country{Singapore}
}
\email{hoonwei.lim@ncs.com.sg}

\author{Jin Song Dong}
\affiliation{%
  \institution{National University of Singapore}
  \city{Singapore}
  \country{Singapore}
}
\email{dcsdjs@nus.edu.sg}

\author{Jiaheng Zhang}
\affiliation{%
  \institution{National University of Singapore}
  \city{Singapore}
  \country{Singapore}
}
\email{jhzhang@nus.edu.sg}

\thanks{\textsuperscript{\#} Corresponding Author.} 

\settopmatter{
authorsperrow=4,
}
\renewcommand{\shortauthors}{Ming Xu et al.}


\begin{abstract} 
Security Information and Event Management (SIEM) systems make it possible for detecting intrusion anomalies in real-time manner by their applied security rules. 
However, the heterogeneity of vendor-specific rules (e.g., Splunk SPL, Microsoft KQL, IBM AQL, Google YARA-L, and RSA ESA) makes cross-platform rule reuse extremely difficult, requiring deep domain knowledge for reliable conversion.  
As a result, an autonomous and accurate rule conversion framework can significantly lead to effort savings, 
preserving the value of existing rules. 
In this paper, we propose \name, an agentic SIEM-rule conversion approach. Using \name, the security professionals do not need to distill the source rules' logic and re-map it to target vendors, instead, they provide the source rules, the documentation of the target rules and \name can purposely convert to the target vendors without more intervention.
To achieve this, \name is equipped with conversion intermediate representation that aligns core detection logic into vendor-neutral layer, agentic RAG pipeline that retrieves authoritative official vendor documentation to address the convention/schema mismatches, and Python-based consistency check that running both source and target rules in controlled test environments to mitigate subtle semantic drifts. 
We present a comprehensive evaluation of \name ranging from textual alignment 
and the execution success, showcasing \name can convert rules with higher fidelity, outperforming the baseline LLM models by 15\% averagely.
Finally, we perform case studies and interview with our industry collaborators in Singtel Singapore,
which showcases that \name can significantly save expert's time 
on understanding cross-SIEM's documentation and remapping logic.  
\end{abstract}

\ccsdesc[500]{Security and privacy}
\ccsdesc[500]{Security and privacy~Software and application security} 

\keywords{Rule-based Intrusion Detection, Agentic AI, AIOps}

\maketitle

\section{Introduction} 
Every year, approximately trillions of web intrusion attempts are made globally~\cite{global-threat}, posing severe threats to web-facing and cloud infrastructures. 
Security Operation Centers (SOCs) serve as frontline defense, tasked with continuously monitoring and responding to these threats in real time. 
While numerous studies show that neural-network-based provenance graphs~\cite{DBLP:journals/corr/abs-2308-05034:Kairos, Flash, DBLP:conf/sp/Hassan0M20:TTP-provance} achieve high detection accuracy, however, their widespread adoption is hindered by computational complexity, expensive training costs, and a lack of interpretability. 
In contrast, most SOCs rely on Security Information and Event Management (SIEM) platforms~\cite{DBLP:journals/corr/abs-2311-10197:SIME-rule-evasion, DBLP:journals/ieeesp/BhattMZ14:SIEM}, which aggregate, filters and alert web and system logs to provide situational awareness in real-time.   
Prominent SIEM platforms including Splunk~\cite{splunk}, Microsoft Sentinel~\cite{microsoftMicrosoftSentinel}, 
Google Chronicle~\cite{Google-chronicle}, 
typically detect malicious activities (e.g., brute-force login) via the inherent SIEM-rules executed by their underlying analysis engines.

However, the effectiveness of SIEM platforms is tightly coupled with rules, which are a SQL-like constraint, that drive their analysis engines. 
In practice, organizations often undergo platform migration~\cite{SIEM-migration-one}, mergers~\cite{SIEM-migration-one} and acquisitions, or operate multiple SIEM vendors in parallel. Such transitions render existing rules incompatible, as each SIEM adopts its own proprietary rules with distinct syntax, semantics, and schema requirements. 
Rule conversion can be performed manually by security experts, which are slow and imposes a heavy workload.  Besides, organizations can rely on static vendor-provided tools, such as Microsoft-provided SPL2KQL~\cite{SPL2KQL}, which supports a narrow conversion from Splunk to Microsoft Sentinel, leaving other platforms unsupported. Recently, they can also involve leveraging the large language models (LLMs) through prompt engineering for rule converter; however, this typically yield a poor accuracy and lacks vendor-specific correctness due to the less exposure of SIEM's corpora of an LLM. 
These shortcomings call for a scalable, vendor-neutral, and reliable SIEM-rule conversion framework that retains existing rule value and eases SOC workloads.

The recent surge in the capabilities of agentic LLM-based workflows in code translation~\cite{DBLP:conf/ndss/LiWLSK25:CtoRust}, SQL translation~\cite{SQL-translation,DBLP:conf/aidm/NgomK24:SQL-translation} and software comprehension~\cite{DBLP:conf/emnlp/FengGTDFGS0LJZ20:codebert, wei2025swerladvancingllmreasoning:software-understanding, liu2025purpcodereasoningsafercode} with generative models like the GPT series, presents a paradigm-shifting opportunity for security rule conversion.  
Unfortunately, compared to SQL translation, SIEM-specific rule conversion is significantly more challenging due to the subtle and nuanced nature. 
For example, SQL has a well-defined standard that most cases follow with minor syntactic differences, while SIEM rule languages such as Splunk SPL, Microsoft KQL, IBM AQL, Google YARA-L, RSA ESA lack a unified specification. 
Each introduces proprietary operators, platform-specific constructs 
and unique data semantics tied to the vendor’s logging pipeline.
This heterogeneity makes rules syntactically different and semantically divergent, where equivalent operators, aggregations, or correlation logic may behave inconsistently across systems. 
For example, Microsoft KQL's \texttt{project} operator maps to Splunk SPL's \texttt{fields}.
As a result, rule conversion requires deeper reasoning about execution semantics and domain-specific understandings, going far beyond the scope of SQL dialect translation with the challenges below:    

\begin{itemize}[fullwidth,itemindent=0em]
     \item \textbf{Heterogeneous rule syntax across SIEM platforms.} 
     Each platform, such as Splunk, Microsoft Sentinel, IBM QRadar, Google Chronicle and RSA NetWitness, adopts proprietary query languages, event models, and detection constructs, lacking standardized representations. As a result, direct rule migration is error-prone, with subtle differences in field naming, operators, and aggregation semantics leading to functional inconsistencies. Traditional static or general-LLM-agent-based translation approaches often fail to capture these nuances, leaving gaps in detection coverage. This motivates the design for a conversion intermediate representation (IR)
     that aligns core detection logic into vendor-neutral layer, ensuring consistency while enabling systematic transformation into diverse target rule formats.  
     \item \textbf{Guaranteeing completeness and vendor-specific correctness:} LLMs are trained on general-purpose code (e.g., SQL, Python) and have limited exposure to SIEM-rule languages. Consequently, naive translations often exhibit structural omissions (e.g., missing mandatory sections/fields), syntax errors (e.g., invalid operators, clause ordering, parameter shapes), and incorrect keyword usage (e.g., misapplied macros/functions, wrong field paths or namespaces) when targeting different SIEM vendors. 
     However, the core challenge here is not merely a matter of ``syntactic translation'' that enforces equivalence with the source rule, but in performing semantic mapping and knowledge supplementation to ensure the converted rule is complete and conforms to the vendor's grammar and conventions.  
     To address this knowledge gap, we employ an agentic retrieval-augmented generation (Agentic-RAG) pipeline that dynamically retrieves authoritative official vendor documentation during generation, using it to enforce template compliance, validate allowed keywords/operators, normalize field references, and fill in required metadata-thereby improving format completeness and syntax correctness of the converted rules. 
    \item \textbf{Verifying functional consistency through executable testing:} 
    A further challenge arises in verifying whether the translated rules are functionally consistent with the source rules. 
    Manually validating detection behavior against real logs is costly and slow, particularly when dealing with nested conditions, custom macros, or event aggregation. To overcome this, we propose generating Python-based executable code blocks that simulate rule execution. By synthesizing representative log data and running both source and target rules in controlled test environments, the system can check whether they yield equivalent outputs. This automated validation loop provides a concrete measure of correctness and supports iterative optimization of translated rules. 
\end{itemize}

In this paper, we address the above problems by proposing \name, an agentic framework for autonomous rule conversion between heterogeneous SIEM vendors.  
\name distills the rule's core logic into a vendor-agnostic layer of conversion intermediate representation (IR), which can be converted into the target rule draft by LLM's reasoning capabilities.  
Then, two autonomous reflection agents including Agentic-RAG pipeline and Python-based consistency check are introduced to dynamically correct the draft rules, ensuring faithful conversion between keyword remapping, logic preserving, and functional equivalence. 
Running \name upon 1,492 pairs of rule conversion across five mainstream SIEM platforms, results show that \name consistently improves logic alignment, semantic validity, and execution success, outperforming general LLMs by around 15\% in similarity alignment. 
Finally, we perform case studies with our industry partners, and show that \name can greatly free SOC professionals from the burden of manually piecing together the cross-SIEM's documentation, and logic equivalence re-mapping during our development in operational environments. 

\noindent\textbf{Contributions.} Our contributions are summarized below.
\begin{itemize}[left=0em]
    \item We propose \name, the first framework for efficient cross-SIEM rule conversion. We systematically analyze rules from multiple SIEM systems and derive a conversion Intermediate Representation (IR). In addition, we introduce two agentic reflection mechanisms that enable \name to dynamically refine structural, syntactic, and semantic nuances during conversion.
    \item We conduct a comprehensive evaluation of \name, using models of GPT-5, DeepSeek-V3, LLaMa-3. \name consistently outperforms general baselines across all models, SIEM platforms and evaluation metrics.
    \item We uncover several empirical domain-specific rule conversion case studies in real-world deployment, which demystify the underlying domain and explain why \name achieves superior performance. 
\end{itemize}

\noindent 
We release the source codes~\footnote{\url{https://github.com/LLM4SOC-Topic/ARuleCon}} for community development, while the prototype is being commercialized by our industry partners.

\lstdefinelanguage{yaml}{
    keywords={true,false,null,y,n},
    keywordstyle=\color{red}\bfseries,
    basicstyle=\ttfamily,
    sensitive=false,
    comment=[l]{\#},
    morecomment=[s]{/*}{*/},
    commentstyle=\color{gray}\ttfamily,
    stringstyle=\color{orange}\ttfamily,
    moredelim=[l][\color{green}]{---},
    moredelim=[l][\color{green}]{...},
    moredelim=**[is][\color{cyan}]{§}{§},
    moredelim=**[is][\color{teal}]{@@}{@@},      
    moredelim=**[is][\color{magenta}]{★}{★}     
}
\lstset{ %
  language=yaml,                
  basicstyle=\ttfamily\footnotesize, 
  numbers=none,                   
  numberstyle=\tiny\color{gray},  
  stepnumber=1,                   
  numbersep=5pt,                  
  backgroundcolor=\color{yellow!10},  
  showspaces=false,               
  showstringspaces=false,         
  showtabs=false,                 
  frame=none,                   
  rulecolor=\color{black},        
  tabsize=2,                      
  captionpos=b,                   
  breaklines=true,                
  breakatwhitespace=false,        
  title=\lstname,                 
  keywordstyle=\color{red},      
  commentstyle=\color{dkgreen},   
  stringstyle=\color{mauve},      
  escapeinside={\%*}{*)},         
  morekeywords={*,...},           
  belowskip=0pt,
  aboveskip=0pt
}

\section{Background and Motivation}

\subsection{A Tour of SIEM and Its Applied Rules}  
Rule-based anomaly detection searches the predefined rule logic via the Security Information and Event Management (SIEM)~\cite{DBLP:journals/ieeesp/BhattMZ14:SIEM, DBLP:journals/corr/abs-2311-10197:SIME-rule-evasion} platforms' analysis engine, which provides APIs to pass the rule queries to the underlying parsing and indexing engine, returning detection results such as matched logs (i.e.,SQL injection, Cross-site Scripting, or Remote File Inclusion). 
At the core, these rules perform \texttt{pattern matching} on fields (e.g., IP addresses, status codes, command strings) and leverage \texttt{statistical thresholds} (e.g., repeated login failures within a fixed time window) to capture abnormal behaviors. Advanced rules perform \texttt{temporal correlation} across multiple events, linking together seemingly benign actions into a potential attack sequence, while contextual enrichment (e.g., cross-checking against threat intelligence feeds or user behavior baselines) further enhances accuracy. Unlike black-box machine learning approaches, rule-based detection provides interpretable and instant detections, allowing analysts to quickly understand the rationale and respond effectively in real time.  
As shown in Table~\ref{tab:rules}, query-based languages such as Splunk’s SPL, Microsoft Sentinel’s KQL, and IBM QRadar’s AQL express detections as dataflow pipelines over tabular events—combining filters, projections, joins, aggregations, and time binning to suspicious patterns. Pattern-matching languages such as Google Chronicle’s YARA-L adopt a more declarative style, describing conditions and (optionally) ordered event sequences over a unified data model, which favors concise, reusable matching logic for threat hunting. RSA NetWitness ESA encodes temporal patterns and stateful correlations directly over event streams.

\begin{table}[htbp]
\vspace{-3.2mm}
\centering
\setlength{\abovecaptionskip}{0pt}
\setlength{\belowcaptionskip}{0pt}
\caption{Mainstream Business SIEM-rules.}  
\label{tab:rules}
\renewcommand\tabcolsep{4.1pt} 
\footnotesize
\begin{tabular}{ccccc} 
\toprule[\thickline]
SIEM-Rules              &  &  Platforms            &  & Type          \\ \cmidrule{1-1} \cmidrule{3-3} \cmidrule{5-5}  
Search Processing Language (SPL) &  & Splunk             & & query-based      \\ \cmidrule{1-1} \cmidrule{3-3} \cmidrule{5-5} 
Kusto Query Language (KQL)       &  & Microsoft Sentinel &  & query-based         \\ \cmidrule{1-1} \cmidrule{3-3} \cmidrule{5-5}    
Ariel Query Language (AQL)       &  & IBM QRadar         &  & query-based       \\ \cmidrule{1-1} \cmidrule{3-3} \cmidrule{5-5} 
YARA-L                           &  & Google Chronicle   & & pattern-based   \\ \cmidrule{1-1} \cmidrule{3-3} \cmidrule{5-5}
Event Stream Analysis (ESA)      &  & RSA NetWitness     & & pattern-based  \\ \bottomrule[\thickline]
\end{tabular}
\vspace{-3mm}
\end{table}

\begin{figure}[htbp]
\setlength{\abovecaptionskip}{0pt}
\setlength{\belowcaptionskip}{0pt}
    \centering
    \scalebox{1}{
\includegraphics[width=0.99\linewidth]{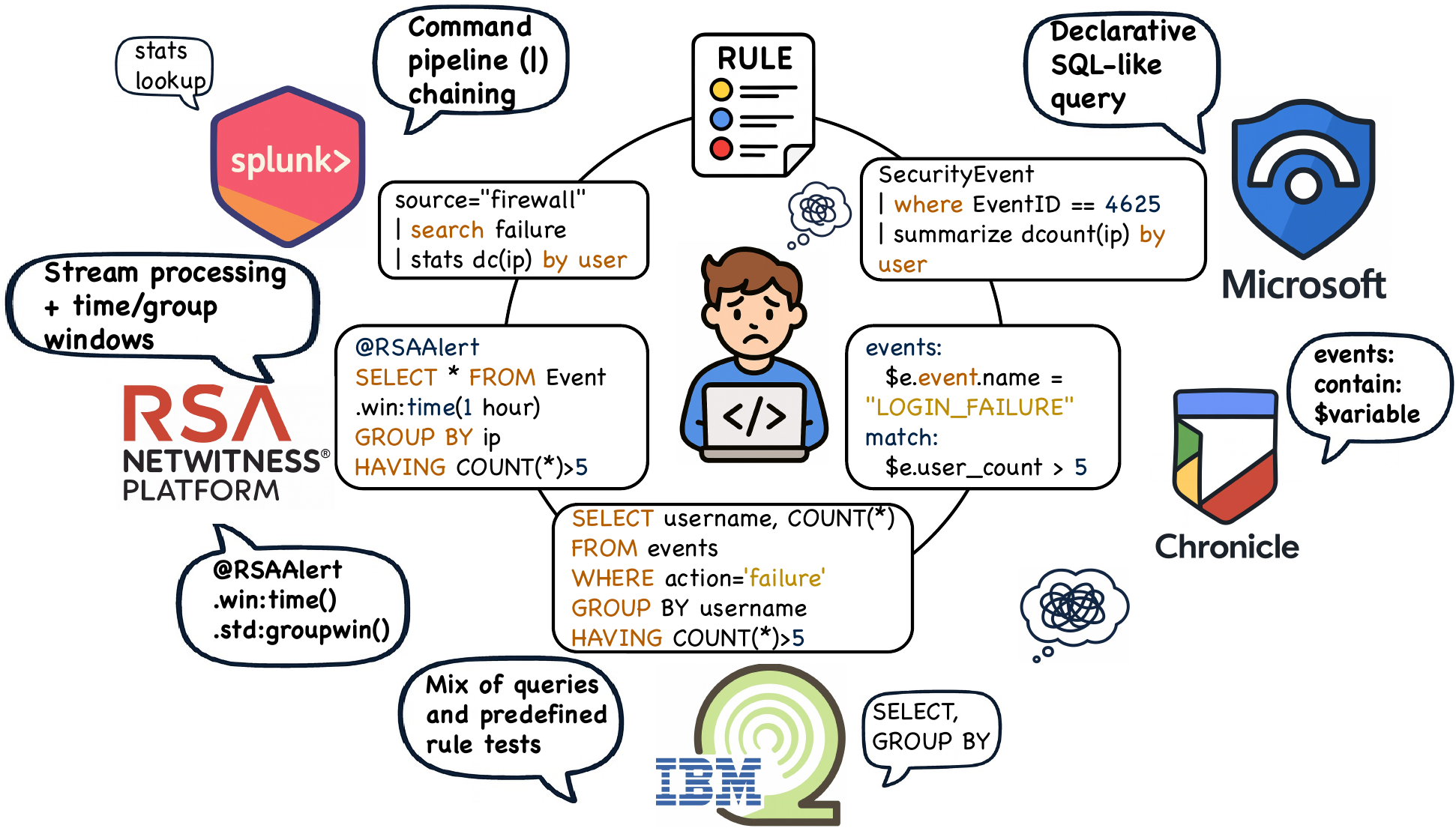}}
    \caption{Motivation Scenarios~\cite{SIEM-migration-one, SIEM-migration-two}: industrial SOCs are plagued by numerous rule query languages that share the backbones of first order logic.  
    }  
    \label{fig:motivation}
    \vspace{-6mm}
\end{figure}

\noindent\textbf{Motivating Principles.} 
Despite the their rules' syntactic variety, these rule configurations share a common backbone: (i) predicates over normalized fields, (ii) temporal scoping, (iii) aggregation with thresholding, (iv) optional context enrichment, and (v) analyst-readable logic. The underlying logic of most rules can typically be expressed in first-order logic, which provides a common foundation for rule transformation. 
However, cross-platform realization of ostensibly similar logic is sensitive to differences in schema conventions, temporal semantics, string-processing semantics, operator sets, and execution models, as shown in Figure~\ref{fig:motivation}.
Variations in normalization frameworks, windowing formalisms, evaluation order, and state management can yield non-isomorphic behavior under the same detection intent, with measurable impact on coverage, cadence, and latency. 
As operational environments typically evolve and migrate, portability and semantic fidelity become first concerns.   
Organizations frequently undergo mergers, budget adjustments, or technology upgrades that necessitate switching SIEM platforms or running multiple systems in parallel, making it essential to preserve previously crafted rules as a core SOC asset to maintain consistency and effectiveness.

\vspace{-2mm}
\subsection{Operational Observations} 
We empirically test conversion cases, 
observing that heterogeneous rule conversion is not a line-by-line or variable-to-variable mapping. 
For example, in Chronicle YARA-L, the \texttt{aggregate} operator is overloaded and can simultaneously express grouping, filtering, and thresholding in a single clause like:  
\begin{lstlisting}[caption={This compact rule states that login failure events should be grouped by \texttt{src\_ip}, restricted to a 30-minute window, and only groups with more than five failures should be retained.}, label={lst:source-rule}]
events:
  filter: event.type = "login_failure"
  §aggregate: count() > 5§ by §src_ip§ §within 30m§
\end{lstlisting}

\noindent A general-LLM-based conversion often attempts to map \texttt{aggregate} literally into QRadar AQL. Since AQL has no single operator with equivalent semantics, the model typically generates invalid or incomplete queries—for example, producing a \texttt{GROUP BY} without the threshold filter, or misplacing the condition into a simple \texttt{WHERE} clause, which changes the meaning of the rule entirely:
\begin{lstlisting}[caption={The threshold \texttt{COUNT(*) > 5} is wrongly placed in the \texttt{WHERE} clause, which is syntactically invalid in AQL, and the 30-minute window constraint is ignored completely.}, label={lst:wrong-case}]
SELECT src_ip, COUNT(*) 
FROM events 
WHERE event_type = 'login_failure' ★AND COUNT(*) > 5★
GROUP BY src_ip, ★log_source_time★
\end{lstlisting}

\noindent This motivates the design of semantic decomposition into a vendor-agnostic specification 
to yield a faithful conversion.


Second, syntactic operators for the same execution can differ significantly. For example, KQL’s \colorbox{mygray}{\texttt{summarize}} and \colorbox{mygray}{\texttt{project}} have no direct equivalents in SPL, where the corresponding operators are \colorbox{mygray}{\texttt{stats}} and \colorbox{mygray}{\texttt{fields}/\texttt{table}}. This discrepancy makes direct operator mapping by LLMs particularly challenging. This motivates the agentic probe of the vendor-specific documentation, enabling the model to consistently correct the conventions. 


Third, the subtle semantic drift and functional inconsistency somewhat occurs.
For example, in Splunk SPL:
\begin{lstlisting}[caption={This rule explicitly groups login failure events by \texttt{src\_ip}, counts the number of failures for each address, and raises an alert when any single IP exceeds five attempts.}, label={lst:splunk}]
index=auth action=failure
| §stats count by src_ip§
| §where count > 5§
\end{lstlisting}

\noindent During conversion into Chronicle YARA-L, the
translation collapses the semantics of the SPL pipeline into a single \texttt{aggregate} clause: 

\begin{minipage}{0.23\textwidth}
\begin{lstlisting}[caption={\footnotesize{LLM-based Conversion}}, label={lst:YARA-L:incorrect}]
events:
  filter: 
    action = "failure"
  aggregate: 
    ★count() > 5★
\end{lstlisting}
\end{minipage}
\hfill
\begin{minipage}{0.23\textwidth}
\begin{lstlisting}[caption={\footnotesize{Faithful Conversion}}, label={lst:YARA-L:correct}]
events:
  filter: 
    action = "failure"
  aggregate: 
    @@count() > 5 by src_ip@@
\end{lstlisting}
\end{minipage}

\noindent  While syntactically valid, the version~\ref{lst:YARA-L:incorrect} omits the grouping by \texttt{src\_ip}, effectively collapsing all failures into a global count.
This failure motivates the functional consistency check between source and target rules. 
All these cases demonstrate that general LLM workflows/executions are insufficient and must be paired with domain-specific agentic workflows that enable semantic decomposition, targeted corrections, and adaptive reflections~\cite{russell2016artificial}.

\section{\name: Methodology} 

\begin{figure}[htbp]
\setlength{\abovecaptionskip}{0pt}
\setlength{\belowcaptionskip}{0pt}
    \centering
    \scalebox{1}{
\includegraphics[width=0.99\linewidth]{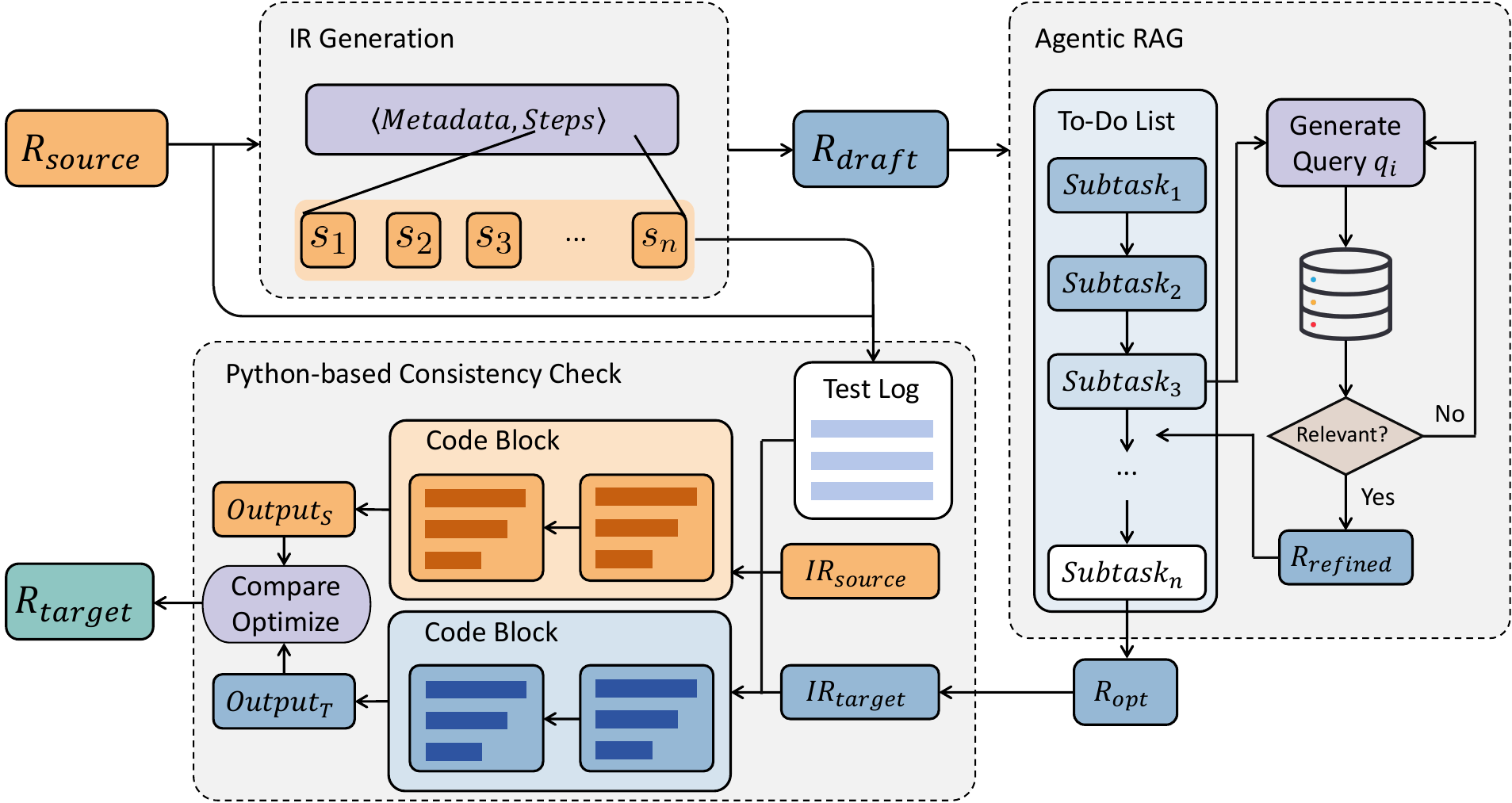}}
    \caption{Overview Pipeline of \name.}  
    \label{fig:overview}
    \vspace{-5mm}
\end{figure}

As shown in Figure~\ref{fig:overview}, the 
workflow of \name is structured into the interconnected generation and reflection stages. 
First, the source rule $R_{\text{source}}$ is normalized into a vendor-agnostic template of intermediate representation (IR). 
Then, the IR is drafted into a target rule $R_{\text{draft}}$ by the generative model, ensuring structural alignment with the grammar of the destination SIEM. 
Third, an Agentic RAG process refines this draft through a to-do list of subtasks, enabling iterative retrieval from vendor documentation.  
Then, a Python-based consistency check validates the semantic equivalence of the source and target rules by compiling their IRs into executable Python blocks, running them over synthetic test logs, and comparing outputs. 


\subsection{IR-driven Source Rule Interpretation}\label{sec:ir}




To handle the heterogeneous rule syntax, our source agent is tasked with parsing the core logic of source rules into a vendor-agnostic template of IR, enabling the model to prioritize semantic logic over syntactic details. 

\noindent\textbf{Intermediate Representation.}  
To normalize SIEM rules into a unified representation, we design the IR as a layered schema that separates fundamental metadata from behavioral logic. This design abstracts away vendor-specific syntactic details while retaining the semantic fidelity of the original rule. Formally, the IR is defined as:
\begin{equation}\nonumber
\footnotesize
\mathit{IR} = \langle M, S \rangle = \langle M, \{ s_1 \circ s_2 \circ \dots \circ s_n \} \rangle
\end{equation}

\noindent where $M$ denotes the set of fixed metadata fields and $S$ represents an ordered list of steps executed sequentially. Specifically:  
\begin{itemize}[fullwidth,itemindent=0em]
  \item $M = \{ \texttt{rule\_name}, \texttt{description}, \texttt{data\_source}, \texttt{event\_type}\}$ provides vendor-independent contextual information, ensuring that the converted rule is properly documented and understood regardless of the underlying SIEM platform.  
  \item $S = \{ s_1 \circ s_2 \circ \dots \circ s_n \}$ captures the ordered logical flow of the rule. Each element $s_i$ in $S$ represents an atomic behavioral unit that the rule
  must preserve during conversion.  
\end{itemize}

\noindent To make this precise, each step $s_i$ is a triplet:  
\begin{equation}\nonumber
\footnotesize
s_i = \langle \mathit{KEYWORD}, \mathit{PARAM}, \mathit{DESCRIPTION} \rangle
\end{equation}


\noindent Where:  

\begin{itemize}[fullwidth,itemindent=0em]
  \item $\mathit{KEYWORD}$: a predefined, vendor-agnostic functional abstraction over SIEM operators. A single \texttt{keyword} may correspond to one or multiple concrete operators across platforms (and conversely, a single vendor operator may decompose into multiple \texttt{keywords}). This keyword decouples semantics from syntax and enables flexible many-to-many mappings across dialects. The predefined set is:  
  \{\texttt{FILTER}, \texttt{EXTRACT}, \texttt{AGGREGATE}, \texttt{OUTPUT}, \texttt{TRANSFORM}, \texttt{RENAME}, \texttt{LOOKUP}, \texttt{BUCKET}, \texttt{JOIN}, \texttt{FILL}, \texttt{APPEND}, \texttt{SORT}, \texttt{DEDUP}, \texttt{APPLY}, \texttt{DEBUG}\}.  

  \item $\mathit{PARAM}$: the core configuration payload of the step, specifying essential arguments such as log sources, filtering predicates, time windows, grouping keys, thresholds, join keys, and other operator parameters. Importantly, we retain all parameters from the source rule in \texttt{param} to minimize information loss, ensuring that even subtle semantics survive the conversion process. 

  \item $\mathit{DESCRIPTION}$: a concise natural-language explanation of the step’s intent. This human-readable description complements the structured fields by allowing LLMs to reason about, summarize, and reorganize rule semantics across heterogeneous dialects. By embedding a semantic gloss at each step, the IR improves alignment during both initial conversion and later reflection.  
\end{itemize}


For example, the YARA-L's overloaded clause (rule~\ref{lst:source-rule}) is decomposed into multiple explicit steps like\footnote{We omit descriptions for space optimization.}:  

\begin{samepage}    
\begin{minted}[fontsize=\small, frame=single, breaklines, bgcolor=lightgray!10]{json}
[
  { "keyword": "filter", "param": "event.type = login_failure" },
  { "keyword": "grouping", "param": "by src_ip" },
  { "keyword": "window", "param": "within 30m" },
  { "keyword": "threshold", "param": "count() > 5" }
]
\end{minted} 
\end{samepage}



\noindent Through this decomposition, the IR bridges the semantic gap between Chronicle’s compact but overloaded \texttt{aggregate} operator (rule~\ref{lst:source-rule}) and the target rule structures. We leverage LLMs as a knowledge-grounded interpreters to parse IR, and leave the detailed prompts in Appendix~\ref{app:prompt-templates}. 

To ensure IR's coverage, all operator keywords across SIEM-vendors are collected manually.
Unlike the IRs designed primarily for rule generation's effectiveness~\cite{DBLP:journals/corr/abs-2511-12224:rulepilot}, our conversion-IR captures cross-SIEM's feasibility, enabling translation across heterogeneous vendors while maintaining semantic compactness. By explicitly capturing both \texttt{keyword} and \texttt{param} for every step, we ensure that no semantic detail from the source rule is lost during the abstraction process. Even subtle thresholds, logical operators, and time windows are preserved, providing a faithful semantic backbone for subsequent translation while maintaining core functionality. 







\subsection{Target Rule Conversion}
From the IR, a draft target rule is first generated to match the syntax and semantics of the destination platform. This draft is then refined through Agentic RAG, where documentation is retrieved and checked for consistency. To ensure functional equivalence, Python executors and synthetic logs are used to compare the outputs of source and target rules. Any mismatches are fed back into an optimization loop for repair and tuning. 
In this way, the process forms a cycle of generation → verification → testing → optimization, creating a closed-loop process during rule conversion.

\subsubsection{Draft Generation}  
The first step in target rule conversion is the automatic drafting of a candidate rule for the destination SIEM platform. This stage leverages the generative capabilities of LLMs to transform the conversion representation $IR$ into a syntactically valid rule guided by the grammar of the target vendor. To ensure both semantic fidelity and structural correctness, we adopt a prompt-engineering strategy that combines chain-of-thought decomposition with vendor-specific instructions.  
We leave the generation mechanism and prompts in Appendix~\ref{app:prompt-templates}.

\subsubsection{Agentic RAG Guidance}\label{sec:rag}
Then, \name incorporates task planning and iterative updates driven by external knowledge. 
The second stage undergoes an \textit{Agentic RAG} reflection, which refines the draft rule through retrieval and reasoning over vendor documentation. 
Unlike standard RAG that performs a single retrieval and passively appends passages to the prompt, Agentic RAG adopts an iterative and adaptive process. It dynamically adjusts its search queries based on the 
retrieved results, refining keywords until useful and precise documentation is obtained. This mimics how a human analyst consults manuals—continuously reformulating searches until the right tutorial or operator explanation is found. Such active guidance is crucial for SIEM rule conversion, where operator semantics are often subtle and not captured by a single retrieval, addressing \emph{field mismatches, and under-specified operator semantics}.  


\noindent\textbf{Design.}   
The refinement of the draft rule $R_{\mathit{draft}}$ is achieved through an agent-driven retrieval loop that decomposes the overall optimization into a sequence of subtasks. Given $R_{\mathit{draft}}$ and the intermediate representation $\mathit{IR}$ (Sec.~\ref{sec:ir}), the system first generates an ordered to-do list
\begin{equation}\nonumber
\footnotesize
\mathcal{T} = \mathrm{GenTodo}(R_{\mathit{draft}},\mathit{IR}) = \langle t_1,\dots,t_m\rangle,
\end{equation}

\noindent where each subtask $t_i$ is defined as
\begin{equation}\nonumber
\footnotesize
t_i = \langle g_i,\, \alpha_i \rangle,\qquad 
\alpha_i:\mathcal{P}(\mathcal{D}_V)\to\{\texttt{accept},\texttt{reject}\}.
\end{equation}
Here $g_i$ denotes the optimization goal of the subtask, such as operator replacement, parameter adjustment, or syntax correction, while $\alpha_i$ is a predicate that evaluates whether a candidate evidence set from the vendor documentation $\mathcal{D}_V$ is sufficient to resolve the goal.

Each subtask is executed in sequence within an isolated context to avoid information overflow. Let the initial state be $R_{\mathit{draft}}^{(0)} = R_{\mathit{draft}}$. For the $i$-th subtask, the agent maintains a working context
\begin{equation}\nonumber
\footnotesize
\mathcal{C}_i=\bigl\{\,t_i,\; R_{\mathit{draft}}^{(i-1)},\; \{q_i^{(j)}\}_{j\ge1},\; \{E_i^{(j)}\}_{j\ge1}\,\bigr\},
\end{equation}
where $q_i^{(j)}$ denotes the $j$-th query generated for task $t_i$, and $E_i^{(j)}$ the retrieved evidence set. The loop proceeds as
\begin{equation}\nonumber
\footnotesize
q_i^{(j)} = \mathrm{GenQuery}\!\bigl(t_i,\, R_{\mathit{draft}}^{(i-1)},\, IR\bigr), \quad
E_i^{(j)} = \mathrm{Retrieve}\!\bigl(q_i^{(j)},\, \mathcal{D}_V\bigr),
\end{equation}
followed by a judgment step
\begin{equation}\nonumber
\footnotesize 
\alpha_i(E_i^{(j)}) \in \{\texttt{accept}, \texttt{reject}\}.
\end{equation}

\noindent If $\alpha_i(E_i^{(j)})=\texttt{reject}$, the agent refines the query
\begin{equation}\nonumber
\footnotesize 
q_i^{(j+1)}=\mathrm{RefineQuery}\!\bigl(q_i^{(j)},\, t_i,\, E_i^{(j)}\bigr),
\end{equation}

\noindent and repeats retrieval until acceptance is achieved. Once a satisfactory evidence set $E_i^{(j^\star)}$ is accepted, it is directly used to optimize the rule. The update step is defined as
\begin{equation}\nonumber
\footnotesize
R_{\mathit{draft}}^{(i)} = \mathrm{Apply}\!\bigl(R_{\mathit{draft}}^{(i-1)},\, E_i^{(j^\star)}\bigr),
\end{equation}
where $E_i^{(j^\star)}$ provides the vendor-grounded guidance for refinement.

Only the updated state $R_{\mathit{draft}}^{(i)}$ is propagated to the next subtask, ensuring that each $t_i$ is processed independently. After all subtasks are executed in sequence, the final refined rule $R_{\mathit{opt}}$ is obtained. 

This design provides a clear formalism: the rule refinement is driven by a sequence of tasks $\mathcal{T}$, each resolved through iterative query generation, retrieval, judgment, and patching, with vendor documentation $\mathcal{D}_V$ serving as the authoritative source of evidence. 
The decomposition of optimization into a to-do list $\mathcal{T}$ prevents uncontrolled context growth: each subtask is executed within its own isolated context, avoiding information/memory overflow and reducing the risk of forgetting earlier evidence.


Our designed agentic RAG loop relies on vendor documentation as the authoritative knowledge source $\mathcal{D}_V$. For each supported SIEM platform, official manuals and rule syntax references are collected. To support fine-grained retrieval, each corpus is pre-processed by extracting its table of contents to build a hierarchical index of supported operators and functions. This structure provides the retrieval agent with coarse-grained entry points to relevant sections.
To improve retrieval precision beyond naive keyword matching, a vendor-specific query set $Q_V=\{q_1,\dots,q_p\}$ is constructed, where each query is annotated with its functional category (e.g., \textit{filtering}, \textit{aggregation}, \textit{temporal window}, \textit{join}, \textit{lookup}). During task execution, the agent dynamically selects or generates queries from $Q_V$ to retrieve documentation fragments most relevant to the current optimization goal. The retrieved passages are then stored as evidence sets $E_i$, which directly support rule updates in the refinement loop.

\subsubsection{Python-based Consistency Check.} 

To fix the subtle semantic drift problems, the final stage of the conversion loop focuses on verifying the semantic consistency between the source rule $Rule_S$ and the converted rule $Rule_T$. To achieve this, we employ a Python-based execution framework that simulates the behavior of SIEM rules under controlled conditions. Test logs $\mathcal{L}$ are generated, consisting of benign events $\mathcal{L}_{\text{normal}}$ and abnormal events $\mathcal{L}_{\text{attack}}$, ensuring that both rules are evaluated under realistic scenarios.

As summarized in Algorithm~\ref{alg:consistency_check}, the converted rule $Rule_T$ is then normalized into a conversion representation $IR_T$, so that both source and target rules are expressed as IRs 
in a comparable format. 
A pipeline executor processes each IR step $s_i \in \mathit{IR}$ sequentially: for every step, Python code is generated and executed,
producing an output 
that becomes the input for the next step. 
In this way, the entire IR functions as a compositional pipeline $f_1 \circ f_2 \circ \dots \circ f_n$ applied to $\mathcal{L}$. After both pipelines are executed, the outputs $O_S$ from $\mathit{IR}_S$ and $O_T$ from $\mathit{IR}_T$ are compared. Any mismatches $\Delta = O_S \triangle O_T$ are analyzed to identify issues such as missing filters, threshold misalignments, or aggregation discrepancies, which then serve as basis for generating optimizations
to refine the target rule.   

\begin{algorithm}
\footnotesize
\caption{Python-based Consistency Check}
\label{alg:consistency_check}
\begin{algorithmic}[1]
\REQUIRE Source rule $Rule_S$, source IR $IR_S$, target rule $Rule_T$
\ENSURE Semantic equivalence score and optimization suggestions

\STATE $\mathcal{L} \gets \text{GenerateTestLogs}(IR_S, Rule_S)$   \textcolor{darkgreen}{/*\texttt{Generate Test Logs} */}

\STATE $IR_T \gets \text{DeriveIR}(Rule_T)$
\textcolor{darkgreen}{/*\texttt{Derive Target IR} */}

\FOR{each $IR \in \{\mathit{IR}_S, \mathit{IR}_T\}$}
    \STATE $data \gets \mathcal{L}$
    \FOR{each step $s_i = \langle \mathit{keyword}, \mathit{param}, \mathit{description} \rangle$ in $IR$}
        \STATE $code \gets \text{GeneratePython}(s_i)$
        \STATE $output \gets \text{Run}(code, data)$
        \STATE $data \gets output$ \COMMENT{feed into next step}
    \ENDFOR
    \STATE $O[IR] \gets data$
\ENDFOR  
\STATE $O_S \gets O[IR_S], \; O_T \gets O[IR_T]$ \textcolor{darkgreen}{/*\texttt{Compare Outputs} */}
\STATE $\mathit{diffs} \gets \text{Compare}(O_S, O_T)$ 
\STATE $\mathit{suggestions} \gets \text{OptimizeTargetRule}(\mathit{diff}s, \mathcal{R}_T)$ \textcolor{darkgreen}{/*\texttt{Optimize Target Rule} */}  
\RETURN $\mathit{suggestions}$
\end{algorithmic}
\vspace{-1mm} 
\end{algorithm}


Take the source SPL rule~\ref{lst:splunk} and the target YARA-L rule~\ref{lst:YARA-L:incorrect} as an example, to verify the equivalence, our system generates executable Python functions below. 
Each function takes a list of dictionaries (representing logs) as input, executes the rule logic, and returns the result.  
The source executor correctly outputs \texttt{['1.2.3.4']}, while the naïve 
target incorrectly returns a global match \texttt{['ALL']}. The mismatch is flagged, prompting the system to add the missing grouping, which yields the corrected target rule.

\begin{minted}[fontsize=\small, frame=single, breaklines, bgcolor=lightgray!10]{python}
from typing import List, Dict
test_log = [
  {"action": "failure", "src_ip": "1.2.3.4"},
  {"action": "failure", "src_ip": "1.2.3.4"},
  ...
  {"action": "failure", "src_ip": "5.6.7.8"}]
def exec_source_rule(logs: List[Dict]) -> List[str]:
    # SPL semantics: count failures per src_ip, keep those > 5
    df = pd.DataFrame(logs)
    d = df[df["action"] == "failure"]
    grp = d.groupby("src_ip").size().reset_index(name="cnt")
    return sorted(grp.loc[grp["cnt"] > 5, "src_ip"].tolist())
def exec_target_rule(logs: List[Dict]) -> List[str]:
    # Naïve YARA-L semantics: global count only (incorrect)
    df = pd.DataFrame(logs)
    d = df[df["action"] == "failure"]
    return ["ALL"] if len(d) > 5 else []
\end{minted}

The Python-based Consistency Check serves as the final safeguard in the conversion pipeline by verifying the intended detection functions. 
Unlike purely syntactic comparison, this stage evaluates rules through actual execution over synthesized test logs $\mathcal{L}$, ensuring that logical operators, thresholds, and aggregations behave consistently across heterogeneous SIEM platforms. 
By modeling each IR step $s_i$ as an executable Python function and propagating intermediate results $\mathit{data}_i$ sequentially, the framework exposes subtle inconsistencies that may not be visible at the textual level, such as missing filters or altered evaluation windows. 
The comparison of outputs $O_S$ and $O_T$ highlights semantic gaps $\Delta$ that guide targeted refinements, 
allowing the system to generate concrete optimization suggestions for $Rule_T$.  

\section{Evaluation}
We quantitatively evaluate the following research questions (RQs): RQ1–\textbf{Conversion Accuracy}
: How effective is \name in converting a source rule into the target rule? 
RQ2–\textbf{Ablation Studies}: How does each component of \name contribute to the overall performance? 
RQ3–\textbf{Efficiency Overhead}: What are the latency and costs during the rule conversion process?

\subsection{Experimental Settings}\label{sec:settings}

\subsubsection{Model and Parameters} 
We build \name upon three state-of-the-art LLMs: GPT-5, DeepSeek-V3 (671B), and LLaMA-3 (405B). 
The DeepSeek-V3 and LLaMa-3 models are downloaded from Hugging Face~\cite{huggingfaceDeepseekaiDeepSeekV3Hugging},\cite{huggingfaceMetallamaLlama31405BHugging}. 
To control generation behavior, all models are configured with a temperature of 0.3 (balancing determinism and flexibility), top-p of 0.9 (ensuring controlled diversity), and a maximum response length of 1024 tokens to prevent excessively long outputs. Parameters used in Agentic RAG and Python-based consistency check are shown in Appendix~\ref{app:parameters}.

\subsubsection{Datasets and Setups}  
Toward evaluating the conversion quality across heterogeneous SIEM rules, we collect datasets from five widely-used platforms. 
All datasets are constructed from their official websites or open-sourced repositories, and each rule entry includes the rule body with a natural language description explaining its purpose and application context. 
The statistics of these datasets are summarized in Table~\ref{tab:datasets}.  
The number of rules varies significantly across vendors, ranging from only a few dozen in IBM QRadar and RSA NetWitness to over one thousand in Splunk. To avoid evaluation bias caused by this imbalance, we randomly sample approximately 100 rules from each dataset when the total size exceeds this threshold. We sequentially use each vendor as the source and convert its rules into the remaining four target vendors, yielding 1,492 conversion pairs across five SIEMs. 

\begin{table}[htbp]
\centering
\setlength{\abovecaptionskip}{0pt}
\setlength{\belowcaptionskip}{0pt}
\caption{Summary of SIEM-rule datasets.}  
\label{tab:datasets}
\renewcommand\tabcolsep{2pt} 
\footnotesize 
\begin{tabular}{ccccl}  
\toprule[\thickline]
SIEM-Rules     & \multicolumn{1}{l}{} & Size (Used) & Time & \multicolumn{1}{c}{Focus/Coverage} \\ \cmidrule{1-1} \cmidrule{3-5} 
\texttt{Splunk SPL}~\cite{splunk-dataset}              &                      & 1725 (100) & 2025       & Cloud, Client-Side       \\ \cmidrule{1-1} \cmidrule{3-5} 
\texttt{Microsoft KQL}~\cite{microsoft-dataset}   &                      & 483 (100)  & 2025       &  ASimDNS, ASimFileEvent  \\ \cmidrule{1-1} \cmidrule{3-5} 
\texttt{IBM AQL}~\cite{ibm-dataset}           &                      & 33   & 2019       & General Security Events            \\ \cmidrule{1-1} \cmidrule{3-5} 
\texttt{Google YARA-L}~\cite{google-dataset}  &                      & 348 (100)  & 2025       & AWS, Google Cloud Platform, Github                   \\ \cmidrule{1-1} \cmidrule{3-5} 
\texttt{RSA ESA}~\cite{rsa-dataset}       &                      & 40   & 2018       & Web and Network Detection     \\ \bottomrule[\thickline]
\end{tabular}
\end{table}

\subsubsection{Metrics}   
We use the structural and lexical alignment between the converted rules and the source rules as an objective accuracy measurement as below. 

\begin{itemize}[fullwidth,itemindent=0em]
    \item \textit{CodeBLEU~\cite{Codebleu}.}  
    To evaluate structural fidelity, we follow prior work and adopt CodeBLEU, a code-oriented extension of BLEU. Instead of comparing surface-level tokens alone, CodeBLEU incorporates abstract syntax tree (AST) matching, data-flow consistency, and weighted n-gram matching to capture both lexical and syntactic similarity. In our setting, rules from different vendors are first converted into Python-equivalent code through a uniform prompting template, and CodeBLEU is then computed between the generated code and the reference code.
    This allows us to assess whether the converted rule preserves the same computational logic as the source rules, independent of vendor-specific syntax.  

    \item \textit{Embedding Similarity.}  
    To complement structural comparison, we compute semantic similarity between rules by embedding them into a continuous vector space using text-embedding-ada-002~\cite{openAI-vector-model}. Cosine similarity is then calculated between the generated rule and the source rule.   
    This approach captures semantic alignment even when the rules differ in keywords or ordering, which is reasonable given that SIEM rules often exhibit dialectal variations but convey the same detection logic.  

    \item \textit{Logic Slot Consistency.}  
    Beyond text or vector similarity, we measure whether the essential logical slots are preserved. Each rule is decomposed into a set of semantic slots, including predicates, temporal windows, aggregation functions, join conditions, and threshold expressions. 
    We extract these slots via regular-expression templates for each SIEM dialect, normalize values (e.g., thresholds and time units), and compute slot-level similarity using token overlap and Jaccard-based matching. 
    The metric compares these slots directly, ignoring surface syntactic differences, and reports the proportion of matched slots between the generated and source rules. 
    This provides a fine-grained view of logical equivalence, ensuring that even when keywords diverge, the underlying detection semantic remains intact.  
\end{itemize}

\vspace{-2mm}
We have tested that the similarity measurements correlate positively with functional equivalence, making them lightweight quantification proxies: In our sampled functionally-equivalent conversions (Appendix~\ref{app:semantic-evaluation}), pairs typically score higher on CodeBLEU (8/10), Embedding Similarity (7/10), and Logic Slot Consistency (10/10). 
To show the functional equivalence between source and target rules, we also employ six semantic-fidelity metrics with an LLM-as-a-judge paradigm in Appendix~\ref{app:semantic-evaluation}, avoiding the syntactically similar yet semantically divergent phenomenon.

\noindent\textbf{Execution Success.} We use Python-based parsers~\cite{yarapython, pyyaml, sqlparse} and open-source validation libraries to test whether the converted rules can be executed upon the target SIEMs. Particularly, we implement grammar-level checks and structural validation routines that verify keyword ordering, clause nesting, and aggregation syntax for formats such as SPL, KQL, and AQL.

\subsubsection{Baseline} 
\vspace{-2mm}
We compare \name's performance against a customized LLM based upon the LLMs of GPT-5, DeepSeek-V3, and LLaMa-3, without the intermediate representation, agentic RAG and Python-based reflection mechanisms. 
The baselines generate rules using the same prompts as those employed by \name (shown in Table~\ref{tab:prompt-template}). 
Both \name and baseline are under the same parameter settings for fair comparison.

\begin{table*}[htbp]
\centering
\setlength{\abovecaptionskip}{0pt}
\setlength{\belowcaptionskip}{0pt}
\caption{Similarity comparison between the converted rules and source rules from \name (AC) and baselines (BL).} 
\label{tab:results}
\renewcommand\tabcolsep{4.2pt}
\footnotesize
\begin{tabular}{l|>{\columncolor{mygray}}c>{\columncolor{mygray}}c>{\columncolor{mygray}}c>{\columncolor{mygray}}c >{\columncolor{mygray}}cc|>{\columncolor{mygray}}cc>{\columncolor{mygray}}cc>{\columncolor{mygray}}cc|>{\columncolor{mygray}}cc>{\columncolor{mygray}}cc>{\columncolor{mygray}}cc}
\toprule[\thickline]
\multirow{2}{*}{Source SIEM $\rightarrow$ Target SIEM}      & \multicolumn{6}{c|}{CodeBLEU ($\uparrow$)}                                                       & \multicolumn{6}{c|}{Embedding Similarity ($\uparrow$)}                                           & \multicolumn{6}{c}{Logic Slot Consistency ($\uparrow$)}                                         \\
                                                  & \multicolumn{2}{c|}{GPT-5}         & \multicolumn{2}{c|}{DeepSeek-V3}          & \multicolumn{2}{c|}{LLaMa-3} & \multicolumn{2}{c|}{GPT-5}         & \multicolumn{2}{c|}{DeepSeek-V3}          & \multicolumn{2}{c|}{LLaMa-3} & \multicolumn{2}{c|}{GPT-5}         & \multicolumn{2}{c|}{DeepSeek-V3}          & \multicolumn{2}{c}{LLaMa-3} \\
                                                  & AC   & \multicolumn{1}{c|}{BL}   & AC   & \multicolumn{1}{c|}{BL}   & AC           & BL          & AC   & \multicolumn{1}{c|}{BL}   & AC   & \multicolumn{1}{c|}{BL}   & AC           & BL          & AC   & \multicolumn{1}{c|}{BL}   & AC   & \multicolumn{1}{c|}{BL}   & AC          & BL          \\ \midrule
Splunk $\rightarrow$ Microsoft Sentinel           & 63.7 & \multicolumn{1}{c|}{58.9} & 60.5 & \multicolumn{1}{c|}{55.4} & 58.6         & 52.9        & 69.3 & \multicolumn{1}{c|}{63.8} & 66.1 & \multicolumn{1}{c|}{60.3} & 62.7         & 57.1        & \cellcolor{mydeepgray} 73.2 & \multicolumn{1}{c|}{67.6} & \cellcolor{mydeepgray} 69.4 & \multicolumn{1}{c|}{63.8} & \cellcolor{mydeepgray} 66.2        & 60.9        \\
Splunk $\rightarrow$ IBM QRadar                   & 63.4 & \multicolumn{1}{c|}{58.2} & 61.1 & \multicolumn{1}{c|}{55.9} & 59.0         & 53.5        & 64.1 & \multicolumn{1}{c|}{58.6} & 62.3 & \multicolumn{1}{c|}{56.9} & 60.4         & 54.8        & \cellcolor{mydeepgray} 65.5 & \multicolumn{1}{c|}{60.2} & \cellcolor{mydeepgray} 63.1 & \multicolumn{1}{c|}{57.7} & \cellcolor{mydeepgray} 60.8        & 55.3        \\
Splunk $\rightarrow$ Google Chronicle             & 63.1 & \multicolumn{1}{c|}{58.0} & 60.2 & \multicolumn{1}{c|}{55.0} & 58.1         & 53.1        & 83.6 & \multicolumn{1}{c|}{77.9} & 80.2 & \multicolumn{1}{c|}{74.7} & 77.5         & 71.8        & \cellcolor{mydeepgray} 68.0 & \multicolumn{1}{c|}{62.6} & \cellcolor{mydeepgray} 65.1 & \multicolumn{1}{c|}{59.6} & \cellcolor{mydeepgray} 62.9        & 57.4        \\
Splunk $\rightarrow$ RSA NetWitness               & 62.8 & \multicolumn{1}{c|}{57.6} & 60.0 & \multicolumn{1}{c|}{55.1} & 58.2         & 53.0        & 73.2 & \multicolumn{1}{c|}{67.9} & 70.4 & \multicolumn{1}{c|}{65.2} & 67.1         & 61.9        & \cellcolor{mydeepgray} 68.4 & \multicolumn{1}{c|}{63.1} & \cellcolor{mydeepgray} 65.7 & \multicolumn{1}{c|}{60.5} & \cellcolor{mydeepgray} 62.6        & 57.2    
\\
\midrule
Microsoft Sentinel $\rightarrow$ Splunk           & 63.7 & \multicolumn{1}{c|}{58.6} & 61.0 & \multicolumn{1}{c|}{55.8} & 58.9         & 53.6        & 65.2 & \multicolumn{1}{c|}{59.8} & 62.7 & \multicolumn{1}{c|}{57.4} & 60.1         & 54.9        & 70.9 & \multicolumn{1}{c|}{65.4} & 67.8 & \multicolumn{1}{c|}{62.3} & 64.2        & 58.8        \\
Microsoft Sentinel $\rightarrow$ IBM QRadar       & 67.4 & \multicolumn{1}{c|}{61.8} & 64.2 & \multicolumn{1}{c|}{58.9} & 61.1         & 55.7        & 54.6 & \multicolumn{1}{c|}{49.4} & 52.9 & \multicolumn{1}{c|}{47.6} & 50.3         & 45.2        & 61.3 & \multicolumn{1}{c|}{56.0} & 59.2 & \multicolumn{1}{c|}{53.7} & 56.5        & 51.1        \\
Microsoft Sentinel $\rightarrow$ Google Chronicle & 67.2 & \multicolumn{1}{c|}{61.6} & 64.0 & \multicolumn{1}{c|}{58.7} & 61.0         & 55.6        & 75.8 & \multicolumn{1}{c|}{70.3} & 73.1 & \multicolumn{1}{c|}{67.7} & 70.2         & 64.9        & 62.1 & \multicolumn{1}{c|}{56.8} & 59.7 & \multicolumn{1}{c|}{54.5} & 57.1        & 51.9        \\
Microsoft Sentinel $\rightarrow$ RSA NetWitness   & 67.9 & \multicolumn{1}{c|}{62.5} & 64.6 & \multicolumn{1}{c|}{59.4} & 61.6         & 56.1        & 79.7 & \multicolumn{1}{c|}{74.2} & 76.5 & \multicolumn{1}{c|}{71.0} & 73.3         & 68.0        & 64.4 & \multicolumn{1}{c|}{59.0} & 61.7 & \multicolumn{1}{c|}{56.3} & 58.6        & 53.7        \\ \midrule
IBM QRadar $\rightarrow$ Splunk                   & 67.9 & \multicolumn{1}{c|}{62.6} & 64.7 & \multicolumn{1}{c|}{59.2} & 61.8         & 56.4        & 70.3 & \multicolumn{1}{c|}{65.1} & 67.2 & \multicolumn{1}{c|}{62.0} & 64.1         & 59.0         & \cellcolor{mytintgray} 52.2 & \multicolumn{1}{c|}{47.1} & \cellcolor{mytintgray} 50.4 & \multicolumn{1}{c|}{45.2} & \cellcolor{mytintgray} 48.1        & 43.0        \\
IBM QRadar $\rightarrow$ Microsoft Sentinel       & 67.5 & \multicolumn{1}{c|}{62.1} & 64.4 & \multicolumn{1}{c|}{59.0} & 61.2         & 56.0        & 72.5 & \multicolumn{1}{c|}{67.3} & 69.3 & \multicolumn{1}{c|}{64.2} & 66.2         & 61.0        & \cellcolor{mytintgray} 57.4 & \multicolumn{1}{c|}{52.3} & \cellcolor{mytintgray} 55.1 & \multicolumn{1}{c|}{50.0} & \cellcolor{mytintgray} 52.6        & 47.5        \\
IBM QRadar $\rightarrow$ Google Chronicle         & 69.1 & \multicolumn{1}{c|}{63.6} & 65.9 & \multicolumn{1}{c|}{60.5} & 62.8         & 57.3        & 82.4 & \multicolumn{1}{c|}{76.8} & 79.3 & \multicolumn{1}{c|}{74.0} & 76.1         & 70.6        & \cellcolor{mytintgray} 56.9 & \multicolumn{1}{c|}{51.8} & \cellcolor{mytintgray} 54.7 & \multicolumn{1}{c|}{49.6} & \cellcolor{mytintgray} 52.1        & 47.1        \\
IBM QRadar $\rightarrow$ RSA NetWitness           & 68.9 & \multicolumn{1}{c|}{63.3} & 65.7 & \multicolumn{1}{c|}{60.2} & 62.6         & 57.1        & 81.9 & \multicolumn{1}{c|}{76.5} & 79.0 & \multicolumn{1}{c|}{73.6} & 75.8         & 70.3        & \cellcolor{mytintgray} 56.6 & \multicolumn{1}{c|}{51.5} & \cellcolor{mytintgray} 54.3 & \multicolumn{1}{c|}{49.3} &  \cellcolor{mytintgray} 51.7        & 46.8        \\ \midrule
Google Chronicle $\rightarrow$ Splunk             & 62.0 & \multicolumn{1}{c|}{56.8} & 59.4 & \multicolumn{1}{c|}{54.1} & 57.2         & 52.1        & 68.5 & \multicolumn{1}{c|}{63.3} & 66.0 & \multicolumn{1}{c|}{60.7} & 63.4         & 58.2        & 57.6 & \multicolumn{1}{c|}{52.4} & 55.1 & \multicolumn{1}{c|}{50.0} & 52.7        & 47.6        \\
Google Chronicle $\rightarrow$ Microsoft Sentinel & 64.6 & \multicolumn{1}{c|}{59.3} & 61.7 & \multicolumn{1}{c|}{56.6} & 59.1         & 54.1        & 64.7 & \multicolumn{1}{c|}{59.6} & 62.0 & \multicolumn{1}{c|}{57.0} & 59.4         & 54.4        & 62.5 & \multicolumn{1}{c|}{57.2} & 60.1 & \multicolumn{1}{c|}{54.9} & 57.4        & 52.3        \\
Google Chronicle $\rightarrow$ IBM QRadar         & 63.2 & \multicolumn{1}{c|}{58.0} & 60.6 & \multicolumn{1}{c|}{55.4} & 58.4         & 53.1        & 63.2 & \multicolumn{1}{c|}{58.1} & 60.8 & \multicolumn{1}{c|}{55.7} & 58.6         & 53.6        & 52.1 & \multicolumn{1}{c|}{47.0} & 50.2 & \multicolumn{1}{c|}{45.2} & 47.9        & 43.0        \\
Google Chronicle $\rightarrow$ RSA NetWitness     & 65.1 & \multicolumn{1}{c|}{59.9} & 62.0 & \multicolumn{1}{c|}{56.9} & 59.6         & 54.4        & 89.6 & \multicolumn{1}{c|}{84.3} & 86.0 & \multicolumn{1}{c|}{80.9} & 83.4         & 78.1        & \cellcolor{mydeepgray} 75.2 & \multicolumn{1}{c|}{69.8} & \cellcolor{mydeepgray} 72.5 & \multicolumn{1}{c|}{67.2}  & \cellcolor{mydeepgray} 69.7        & 64.4        \\ \midrule
RSA NetWitness $\rightarrow$ Splunk               & 69.7 & \multicolumn{1}{c|}{64.2} & 66.3 & \multicolumn{1}{c|}{61.1} & 63.1         & 57.7        & 63.0 & \multicolumn{1}{c|}{57.9} & 60.4 & \multicolumn{1}{c|}{55.4} & 57.8         & 52.6        & \cellcolor{mydeepgray} 74.1 & \multicolumn{1}{c|}{68.8} & \cellcolor{mydeepgray} 71.3 & \multicolumn{1}{c|}{66.1} & \cellcolor{mydeepgray} 68.6        & 63.4        \\
RSA NetWitness $\rightarrow$ Microsoft Sentinel   & 68.3 & \multicolumn{1}{c|}{62.8} & 65.1 & \multicolumn{1}{c|}{59.7} & 62.0         & 56.6        & 62.3 & \multicolumn{1}{c|}{57.3} & 59.8 & \multicolumn{1}{c|}{54.7} & 57.0         & 52.0        & \cellcolor{mydeepgray} 72.3 & \multicolumn{1}{c|}{67.0} & \cellcolor{mydeepgray} 69.6 & \multicolumn{1}{c|}{64.3} & \cellcolor{mydeepgray} 66.7        & 61.6        \\
RSA NetWitness $\rightarrow$ IBM QRadar           & 71.5 & \multicolumn{1}{c|}{66.0} & 68.2 & \multicolumn{1}{c|}{62.9} & 65.3         & 60.0        & 59.9 & \multicolumn{1}{c|}{54.7} & 57.6 & \multicolumn{1}{c|}{52.4} & 55.1         & 50.1        & \cellcolor{mydeepgray} 81.1 & \multicolumn{1}{c|}{75.8} &\cellcolor{mydeepgray} 78.2 & \multicolumn{1}{c|}{72.9} & \cellcolor{mydeepgray} 75.4        & 70.2        \\
RSA NetWitness $\rightarrow$ Google Chronicle     & 67.2 & \multicolumn{1}{c|}{61.7} & 64.1 & \multicolumn{1}{c|}{58.8} & 61.0         & 55.9        & 64.1 & \multicolumn{1}{c|}{59.0} & 61.4 & \multicolumn{1}{c|}{56.4} & 58.8         & 53.8        &  \cellcolor{mydeepgray} 78.9 & \multicolumn{1}{c|}{73.4} & \cellcolor{mydeepgray} 75.9 & \multicolumn{1}{c|}{70.6} &    \cellcolor{mydeepgray} 73.1      & 67.8        \\ \bottomrule[\thickline]
\end{tabular}
\end{table*}

\subsection{Experimental Results}

\subsubsection{RQ1-Accuracy} 
We present our \textit{quantifiable accuracy assessment} in Table~\ref{tab:results}.
\name (AC) consistently outperforms the baselines (BL) across all models, SIEM platforms, and evaluation metrics. Overall, GPT achieves the strongest absolute performance, while DeepSeek and LLaMA also benefit substantially from \name, confirming that the improvements are model-agnostic. Conversions involving Splunk and RSA NetWitness generally yield higher logic slot similarity, whereas IBM QRadar conversions are relatively more challenging when preserving their logic slot consistency. This is perhaps due to QRadar's stricter separation of stateless vs. stateful tests, complex condition ordering, and performance penalties associated with regex parsing.  

On aggregate, \name delivers consistent gains across all three metrics. For GPT, the average improvement over BL is +9.1\% in CodeBLEU, +10.3\% in embedding similarity, and +11.6\% in logic slot consistency. DeepSeek shows gains of +8.4\%, +11.1\%, and +13.0\% on the same metrics, while LLaMA improves by +9.7\%, +10.8\%, and +12.1\% respectively. When examining the full range, the relative improvements vary between +5\% and +15\% depending on the conversion direction and metric. These results suggest that \name enhances surface-level similarity, strengthens logic preservation as reflected by the larger margins on embedding similarity and logic slot consistency. We have tested \name can also outperform the code-execution agents (i.e., OpenHands, SWE-Agent), which use generic, tool-heavy workflows, and lack the task-specific iterative refinement for precise vendor-specific grammar adaption/subtle semantic-reformatting, always yielding poor performance.

\noindent\textbf{Failure Cases.} To grasp the underlying conversion boundaries, we analyze the common difficulties like nested query structures, and rules involving temporal windows and thresholds in Appendix~\ref{app:temporal-windows-thresholds}.
We manually check the lower-ranked cases of source-target conversions and reveals that the failure is typically caused by the SIEM-specific characteristics, as detailed below: 

\begin{itemize}[fullwidth,itemindent=0em]
    \item \emph{Stateful vs. stateless rule evaluation in IBM QRadar}.    
    QRadar distinguishes between stateless single-event tests and stateful multi-event correlation. Conversion often fails when stateful conditions (e.g., repeated failures within a window) are flattened into stateless filters in target SIEMs, losing the temporal correlation. 

    \item \emph{Temporal and aggregation mismatches in Microsoft Sentinel and Google Chronicle}.
    Sentinel KQL and Chronicle rules support long-range queries, flexible time binning, and large-scale historical aggregation. These constructs may not be supported in platforms with stricter time windows or limited aggregation functions, causing incomplete or invalid translations.
\end{itemize}  

\noindent\textbf{Execution Success.}  
We report the execution validity of the converted rules across heterogeneous SIEM vendors in Table~\ref{tab:execution-success}, presenting results for \name while omitting baseline results due to their significant poor performance. 
Overall, \name achieves high syntax validity rates, with most conversions reaching above 90\%. In particular, conversions involving Google Chronicle and Splunk tend to achieve near-perfect success (close to 100\%). This is likely because these two platforms not only have greater market adoption, which increases the likelihood that LLMs were exposed to their syntax during training, but also provide clearer and more structured official documentation, enabling models to better capture execution
nuances.  
In contrast, conversions targeting IBM QRadar and RSA NetWitness occasionally fall below full validity, with scores around 92–97\%. 
Their weaker performance can be explained by more complex grammar rules, combined with limited publicly available datasets and less comprehensive documentation, 
which together hinder the LLM’s ability to generalize.
Despite these challenges, the results confirm that \name remains lightweight yet reliable in  syntactic verification.  

\begin{table}[htbp]
\vspace{-3mm}
\footnotesize
\setlength{\abovecaptionskip}{0pt}
\setlength{\belowcaptionskip}{0pt}
\centering
\caption{Execution success rate on the target SIEM Platforms.} 
\label{tab:execution-success}
\renewcommand\tabcolsep{1.6pt} 
\begin{tabular}{llccccccccc}
\toprule[\thickline]
\diagbox{Source Rules}{Target Rules} &  & Splunk &  & \begin{tabular}[c]{@{}l@{}}Microsoft \\ Sentinel\end{tabular} &  & \begin{tabular}[c]{@{}l@{}}IBM \\ QRadar\end{tabular} &  & \begin{tabular}[c]{@{}l@{}}Google \\ Chronicle\end{tabular} &  & \begin{tabular}[c]{@{}l@{}}RSA \\ NetWitness\end{tabular} \\ \cmidrule{1-1} \cmidrule{3-3} \cmidrule{5-5} \cmidrule{7-7} \cmidrule{9-9} \cmidrule{11-11} 
Splunk &  & -- &  & 0.868 &  & 0.774 &  & 1.000 &  & 0.868 \\
\cmidrule{1-1} \cmidrule{3-3} \cmidrule{5-5} \cmidrule{7-7} \cmidrule{9-9} \cmidrule{11-11} 
Microsoft Sentinel &  & 1.000 &  & -- &  & 1.000 &  & 1.000 &  & 0.943 \\
\cmidrule{1-1} \cmidrule{3-3} \cmidrule{5-5} \cmidrule{7-7} \cmidrule{9-9} \cmidrule{11-11} 
IBM QRadar &  & 1.000 &  & 0.970 &  & -- &  & 1.000 &  & 0.939 \\
\cmidrule{1-1} \cmidrule{3-3} \cmidrule{5-5} \cmidrule{7-7} \cmidrule{9-9} \cmidrule{11-11} 
Google Chronicle &  & 1.000 &  & 1.000 &  & 1.000 &  & -- &  & 1.000 \\
\cmidrule{1-1} \cmidrule{3-3} \cmidrule{5-5} \cmidrule{7-7} \cmidrule{9-9} \cmidrule{11-11} 
RSA NetWitness &  & 1.000 &  & 0.925 &  & 1.000 &  & 1.000 &  & -- \\
 
\bottomrule[\thickline]
\end{tabular}
\end{table}

\subsubsection{RQ2-Ablation Study}  
\begin{figure}[htbp]
\setlength{\abovecaptionskip}{0pt}
\setlength{\belowcaptionskip}{0pt}
    \centering
\includegraphics[width=0.95\linewidth]{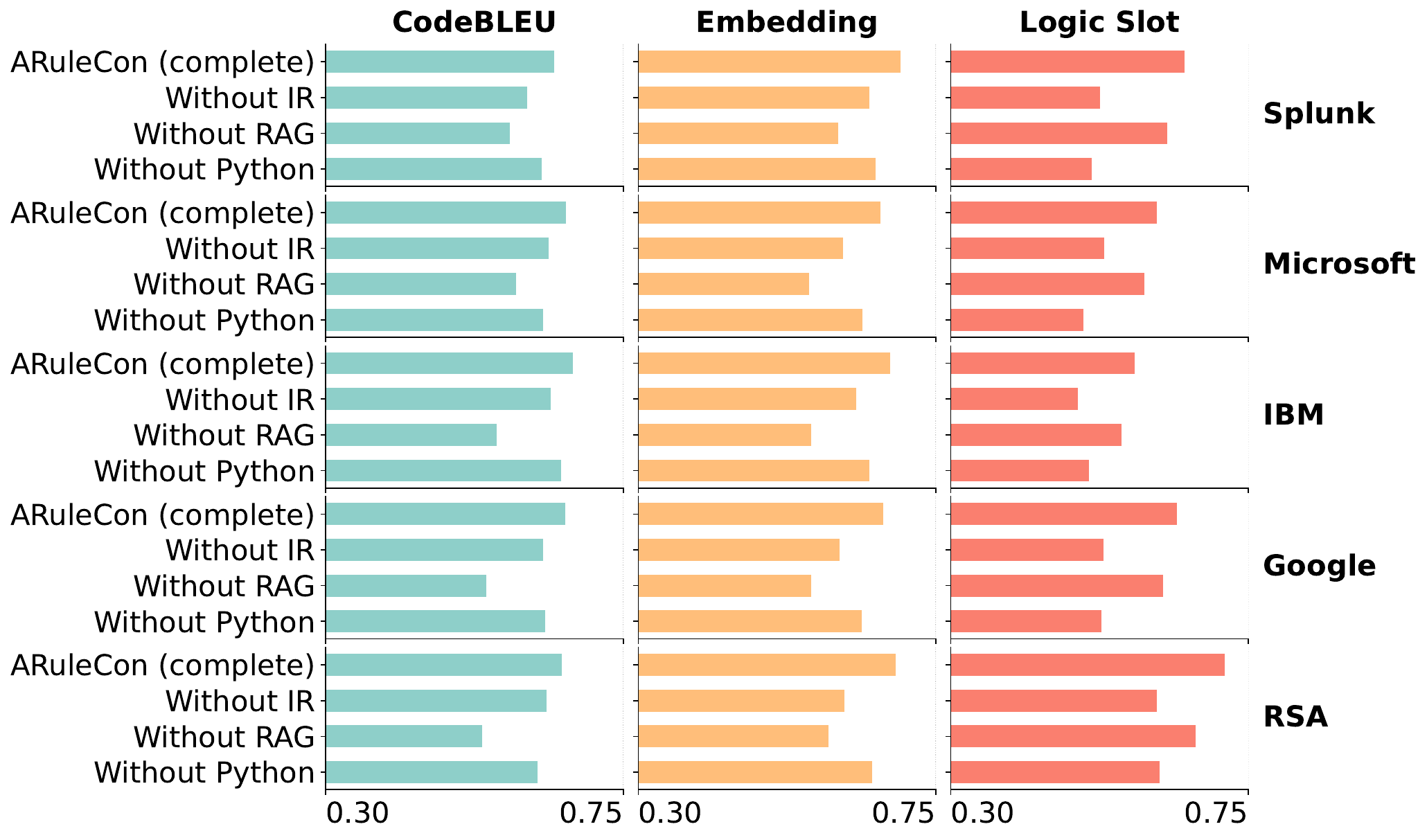}
    \caption{Each component contributes to the performance.}
    \label{fig:ablation}
\vspace{-2mm}
\end{figure}

The results shown in Figure~\ref{fig:ablation} present the ablation study to assess the necessity of each component. 
Without \textbf{IR}, the model loses \textbf{fine-grained logical consistency}, since compact operators (e.g., \texttt{aggregate}, \texttt{tstats}) are harder to decompose without structured steps. Excluding \textbf{Agentic RAG} could lead to lower \textbf{embedding similarity}, as vendor-specific syntax often requires additional documentation support for accurate operator selection. Removing the \textbf{Python-based consistency check} appears to reduce \textbf{logic slot alignment}, as execution validation helps identify subtle issues such as missing \texttt{GROUP BY} or misplaced thresholds.

\vspace{-5mm}
\subsubsection{RQ3-Efficiency} 
We present the computational and economic costs in Table~\ref{tab:cost}, which are derived by running \name and a baseline direct translation on five SIEM rule dialect datasets,
and averaging the results across multiple test cases. 
We observe that \name requires more tokens and computation time than the baseline, mainly due to the
multi-step reflection. 
The main reason could be the optimization steps: 1) the agentic RAG process requires multiple rounds of retrieval to select the most relevant vendor documentation; 2) the generated Python code occasionally contains errors that must be fixed through iterative debugging.
However, as SIEM-rule translation is a high-stake security task, where 15\%-accuracy-gain justifies an additional 100-seconds of computation. The misconverted rules translate into missed detections, which can propagate into downstream incident-response failures. 


\begin{table}[]
\vspace{-3mm}
\footnotesize
\setlength{\abovecaptionskip}{0pt}
\setlength{\belowcaptionskip}{0pt}
\caption{Efficiency and cost of \name and baselines.}
\label{tab:cost}
\renewcommand\tabcolsep{5pt}
\centering
\setlength{\abovecaptionskip}{0pt}
\setlength{\belowcaptionskip}{0pt}
\begin{tabular}{lllllll}
\toprule
Model                        &  &           & \begin{tabular}[c]{@{}l@{}}Prompt\\ Tokens\end{tabular} & \begin{tabular}[c]{@{}l@{}}Output\\ Tokens\end{tabular} & \begin{tabular}[c]{@{}l@{}}Money\\ Cost (USD)\end{tabular} & \begin{tabular}[c]{@{}l@{}}Generation\\ Time (s)\end{tabular} \\ 
\cmidrule{1-1} \cmidrule{3-7} 
\multirow{2}{*}{GPT-5}      &  & \name     & 20,184 & 3,042 & 0.046 & 142 \\
                             &  & Baseline  & 2,137  &   514 & 0.008 &  12 \\ 
\cmidrule{1-1} \cmidrule{3-7} 
\multirow{2}{*}{DeepSeek-V3} &  & \name     & 18,532 & 2,874 & -- & 177 \\
                             &  & Baseline  & 1,954  &   472 & -- &  16 \\ 
\cmidrule{1-1} \cmidrule{3-7} 
\multirow{2}{*}{LLaMA-3}     &  & \name     & 16,947 & 2,601 & -- & 163 \\
                             &  & Baseline  & 1,806  &   438 & -- &  18 \\ 
\bottomrule
\end{tabular}
\vspace{-9mm}
\end{table}

\vspace{-4mm}
\section{Conclusion and Future Works}
In this paper, we propose \name, an agentic rule conversion framework.
By leveraging the intermediate representation and two reflection agents, \name can reliably translate a source SIEM-rule into the target.
Our case study with industry collaborators shows that \name can help industry practices in their operational workflow by lowering the progress of consulting documentation and rewriting queries. 
Future work can collect a larger corpus of paired source-target rules to fine-tune and align the models, further improving accuracy. 

\begin{acks}
This paper is supported by the Minister of Education, Singapore (MOE-T2EP20125-0015), and the NUS-NCS Joint Laboratory for Cyber Security backed up by Singtel Singapore.   
\end{acks}

\IfFileExists{main.bbl}{

}{
\bibliographystyle{ACM-Reference-Format}
\bibliography{my}
}

\appendix

\section{Discussion}

\noindent\textbf{Constraint Translation.}
Agentic AI offers a promising paradigm for constraint translation (e.g., translating the C programming language to Rust~\cite{DBLP:conf/ndss/LiWLSK25:CtoRust}, or SQL translations~\cite{SQL-translation, DBLP:conf/aidm/NgomK24:SQL-translation}), where traditional static approaches often fail to capture subtle semantics and domain-specific dependencies.
ConcoLLMic~\cite{ConcoLLMic} develops a novel approach that uses LLM agents to model symbolic execution: symbolic modeling and constraint solving. 
IntentionTest~\cite{qi2025intention} proposes an intention-constrained approach for generating project-specific test cases. The domain specific constraints are also widely used in SGX-related domains~\cite{wu2021exploring}. 




\noindent\textbf{Takeaways.} 
\name provides insights for code translation, SQL translation, and other domain-specific constraint conversions. 
This methodology suggests that similar layered designs can reduce semantic drift and improve executability in cross-system translation tasks beyond SIEMs.

\section{Prompt Mechanisms in \name}~\label{app:prompt-templates}

\noindent\textbf{IR Generation Mechanism/Prompts.} To ensure reliable extraction of IR from heterogeneous source rules, we leverage LLMs not only as free-form parsers but as knowledge-grounded interpreters. Instead of relying solely on the source rule text, we provide the model with vendor-specific tutorials that are distilled from official SIEM documentation. These tutorials summarize how filtering conditions, aggregation clauses, and temporal windows are typically expressed in each platform, serving as a lightweight guide for the LLM to align its interpretation with vendor conventions.
The detailed tutorials and prompts are shown in our open-sourced code.

\noindent\textbf{Draft Rule Generation Mechanism/Prompts.} The IR is serialized into a structured input containing metadata ($M$) and sequential steps ($S$), which are then injected into a vendor-aware prompt template. The template specifies the SIEM context,  
(e.g., “You are a security analyst specializing in Microsoft Sentinel KQL”), 
enumerates a task list for reasoning (e.g., mapping log sources, applying filters, performing aggregations, validating syntax), and provides few-shot exemplars where available. An example of such a template is shown in Table~\ref{tab:prompt-template}.  
The draft produced at this stage preserves the semantics captured in $\mathcal{R}$ while adhering to the structural requirements of the target vendor. Although not guaranteed to be fully correct, this draft provides a consistent and syntactically valid baseline for subsequent refinement.

\begin{table}[h!]
\footnotesize
\setlength{\abovecaptionskip}{0pt}
\setlength{\belowcaptionskip}{0pt}
\caption{Structure of the Draft Generation Prompt\label{tab:prompt-template}}  
\centering
\begin{tabularx}{\linewidth}{X} 
    \toprule[\thickline]
    \textbf{Vendor-Aware CoT Prompt Template}\\
    \midrule[\thinline]
    \textbf{System Role:} You are a security analyst at a cybersecurity company, specializing in writing and optimizing \textless Target SIEM\textgreater\ rules for threat detection. \\[0.5ex]

    \textbf{Task:} Convert the following conversion Intermediate Representation (IR) into a syntactically valid \textless Target SIEM\textgreater\ rule. \\[0.5ex]

    \textbf{Input (IR):} Metadata $M$ and ordered steps $S = (s_1, s_2, \dots, s_n)$, where each $s_i=\langle \texttt{keyword},\ \texttt{param},\ \texttt{description} \rangle$. \\[0.5ex]

    \textbf{Reasoning Instructions:} Follow chain-of-thought reasoning to:  
    \textcircled{1} Identify relevant event sources from $M$ \\
    \textcircled{2} Map IR \texttt{keywords} to equivalent operators in \textless Target SIEM\textgreater \\
    \textcircled{3} Apply parameters (\texttt{param}) precisely without loss \\
    \textcircled{4} Verify temporal windows, aggregations, and thresholds \\
    \textcircled{5} Construct a rule that is both syntactically correct and semantically faithful \\
    \textcircled{6} Validate against vendor grammar before final output \\[0.5ex]

    \textbf{Output:} A single \textless Target SIEM\textgreater\ detection rule, wrapped in code block format. \\[0.5ex]

    \textbf{Example Input:}  
    \{ "keyword": "AGGREGATE", "param": "count() by src\_ip where action=failure", "description": "Detect source IPs with failed logins" \} \\[0.5ex]

    \textbf{Example Output (KQL):}  
    \texttt{SecurityEvent | where ActionType == "failure" | summarize count() by src\_ip | where count\_ > 5} \\
    \bottomrule[\thickline] 
\end{tabularx}
\end{table}

\section{Evaluation} 
\subsection{Parameters of Components}~\label{app:parameters}
For the Agentic RAG-based reflection component, we construct a vendor-specific retrieval corpus based on the official documentation of Splunk SPL~\cite{splunk-documentation}, Microsoft Sentinel KQL~\cite{microsoft-documentation}, IBM QRadar AQL~\cite{IBM-documentation}, Google Chronicle YARA-L~\cite{google-documentation}, and RSA NetWitness ESA~\cite{rsa-documentation}. 
The documentation is processed through a hierarchical chunking strategy. Each document is first segmented according to its chapters or sections; if a section exceeds a predefined length threshold, it is recursively split into smaller overlapping chunks. Each chunk is then embedded using text-embedding-ada-002~\cite{openAI-vector-model}, producing dense vector representations stored in Chroma that capture semantic similarity across heterogeneous rule descriptions.  
During evaluation, the retrieval agent selects the top-5 most relevant passages for each rule clause and injects them into the prompt, grounding the model’s reflection in authoritative references.

The reflection process in \name is bound to ensure both effectiveness and efficiency. For the Agentic RAG component, when a converted rule exhibits mismatched or incomplete keywords during IR extraction, the model is prompted again with retrieved documentation passages and asked to revise the specific step; this iteration is repeated at most $N=3$ times. For the Python code generation stage, if the generated code fails execution (e.g., syntax errors or runtime mismatches), the system triggers a separate reflection loop in which the model debugs and regenerates the code, with a maximum of $K=5$ attempts.  

\subsection{Semantic-fidelity Evaluation}~\label{app:semantic-evaluation}

\noindent\textbf{Metrics.} The quantifiable similarity assessment offers an objective metric, yet it may yield misleadingly high scores when rules are syntactically similar but semantically incorrect. To mitigate this bias, we adopt the LLM-as-a-judge paradigm, a scalable method for approximating human preferences~\cite{li2024llmsasjudgescomprehensivesurveyllmbased} over the six key evaluation dimensions below.

\begin{itemize}[fullwidth,itemindent=0em]
    \item \textit{Event Scope \& Schema Mapping (E/S).}  
    Whether the source and converted rules operate over the same event sources and fields, including index/table selection, event type alignment, and schema-level field mappings (e.g., \texttt{host} $\leftrightarrow$ \texttt{Computer}, \texttt{src\_ip} $\leftrightarrow$ \texttt{IpAddress}).  

    \item \textit{Predicate \& Boolean Logic (P/B).}  
    Equivalence of filtering conditions, covering operators, negations, case sensitivity, regular expressions, and logical combinations of predicates, such as list membership and range boundaries.  

    \item \textit{Temporal Semantics \& Windows (T/W).}  
    Consistency in time-related constraints, including window size and type (tumbling vs. sliding), alignment anchors, and the placement of temporal conditions before or after aggregation.  

    \item \textit{Aggregation \& Thresholding (A/T).}  
    Alignment of aggregation functions, grouping keys, distinct semantics, threshold comparisons, and whether post-aggregation filtering (e.g., \texttt{having}) is preserved.  

    \item \textit{Correlation \& Joins (C/J).}  
    Correctness of multi-event or multi-source correlation, including join keys, join type, temporal constraints for co-occurrence, and directionality or deduplication strategies.  

    \item \textit{Alert Semantics \& Outcome (A/O).}  
    Preservation of the alerting conditions and outputs, including trigger criteria, deduplication or throttling settings, entities in the result (e.g., IP, user), and severity levels or tagging.  
\end{itemize}

\begin{figure*}[t]
\setlength{\abovecaptionskip}{0pt}
\setlength{\belowcaptionskip}{0pt}
\footnotesize
  \centering
    \subfigure[\texttt{GPT-5}]{\includegraphics[width=0.3\textwidth]{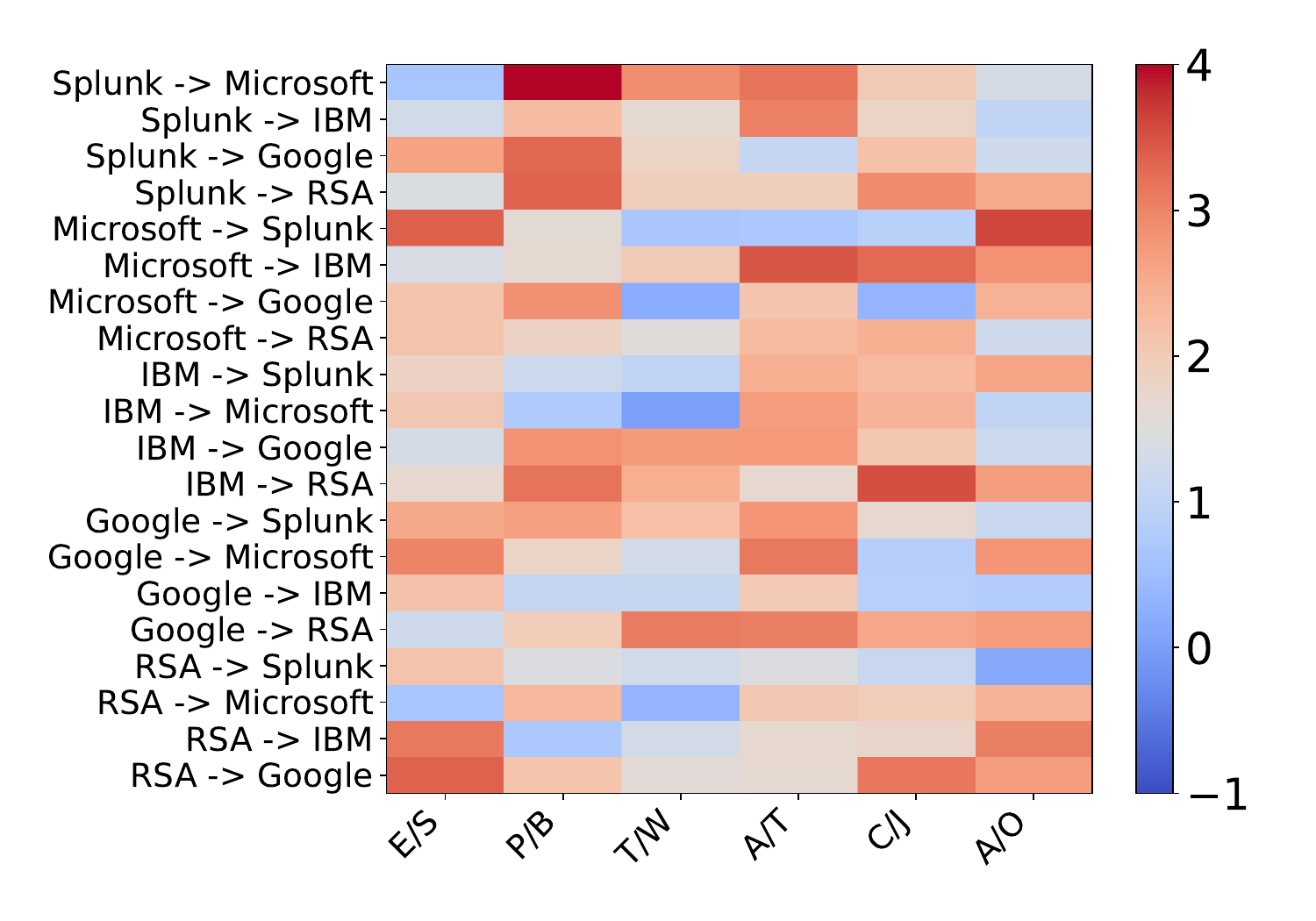}} 
    \subfigure[\texttt{DeepSeek-V3}]{\includegraphics[width=0.3\textwidth]{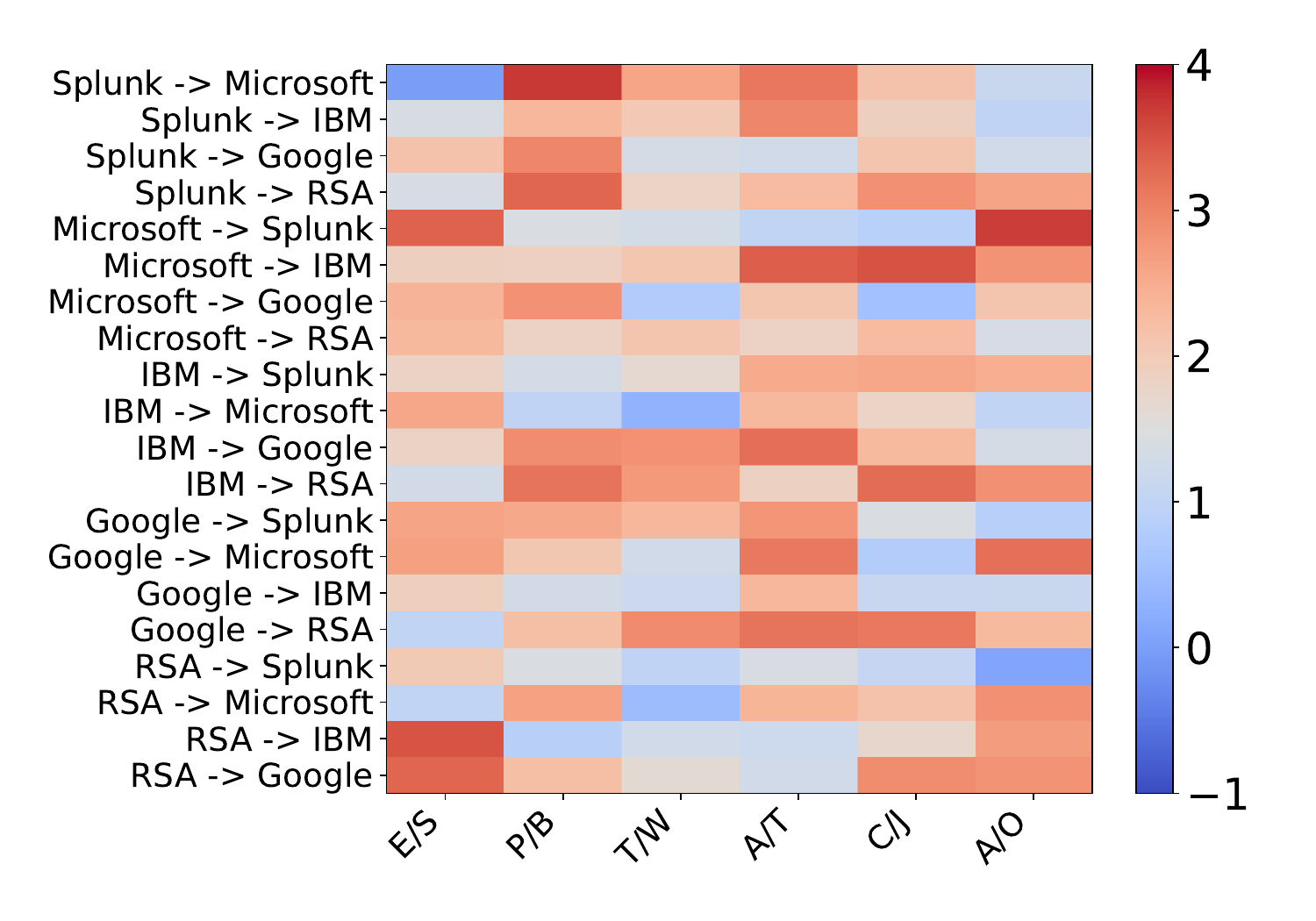}} 
    \subfigure[\texttt{LLaMa-3}]{\includegraphics[width=0.3\textwidth]{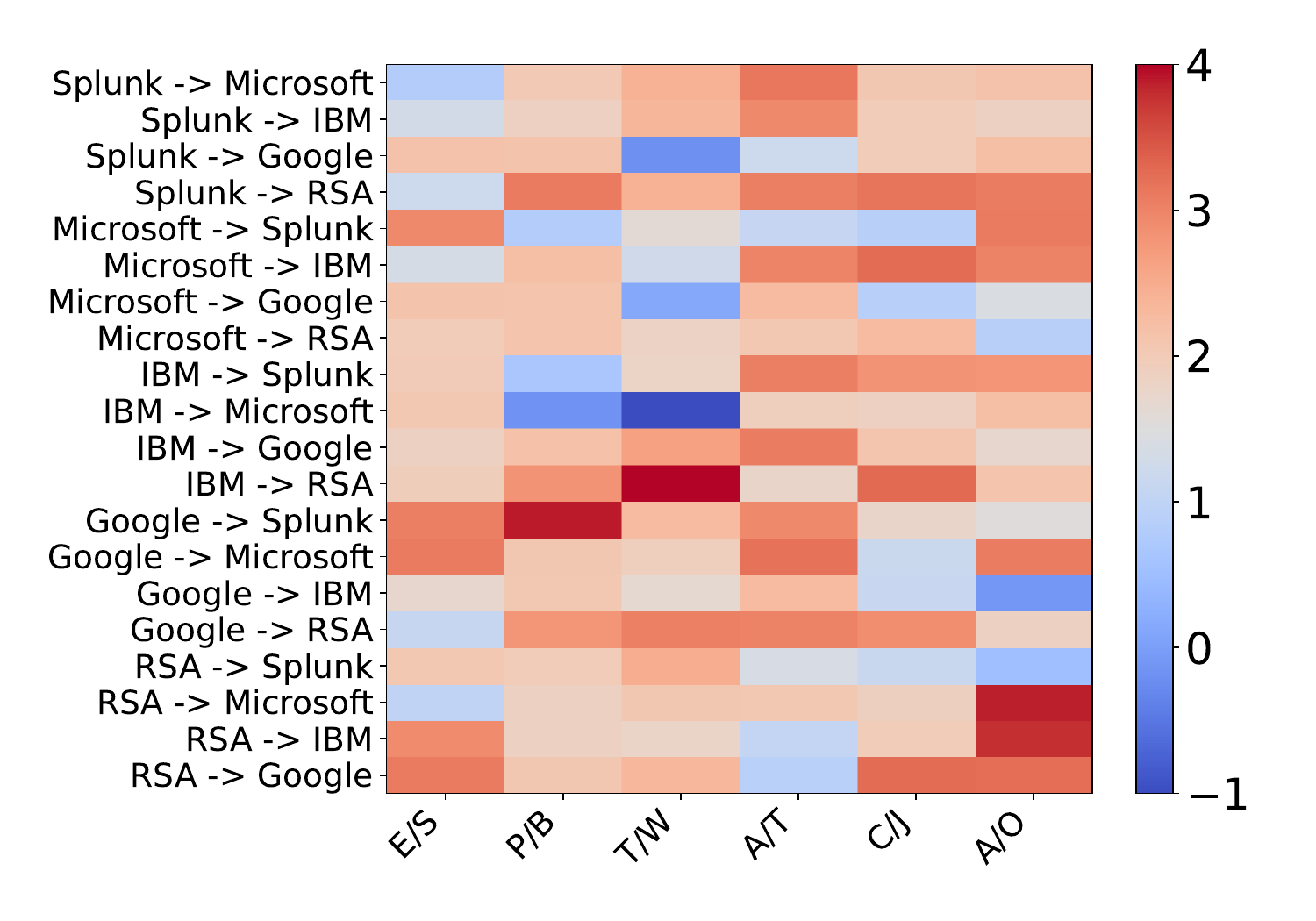}} 
\caption{Fine-grained conversion evaluation on semantic-level: redder regions highlight systematic improvements, and blue cells indicate challenges in schema mapping.} 
\label{fig:radardetails}
\end{figure*}

\begin{figure}[htbp]
\setlength{\abovecaptionskip}{0pt}
\setlength{\belowcaptionskip}{0pt}
    \centering
\includegraphics[width=0.50\linewidth]{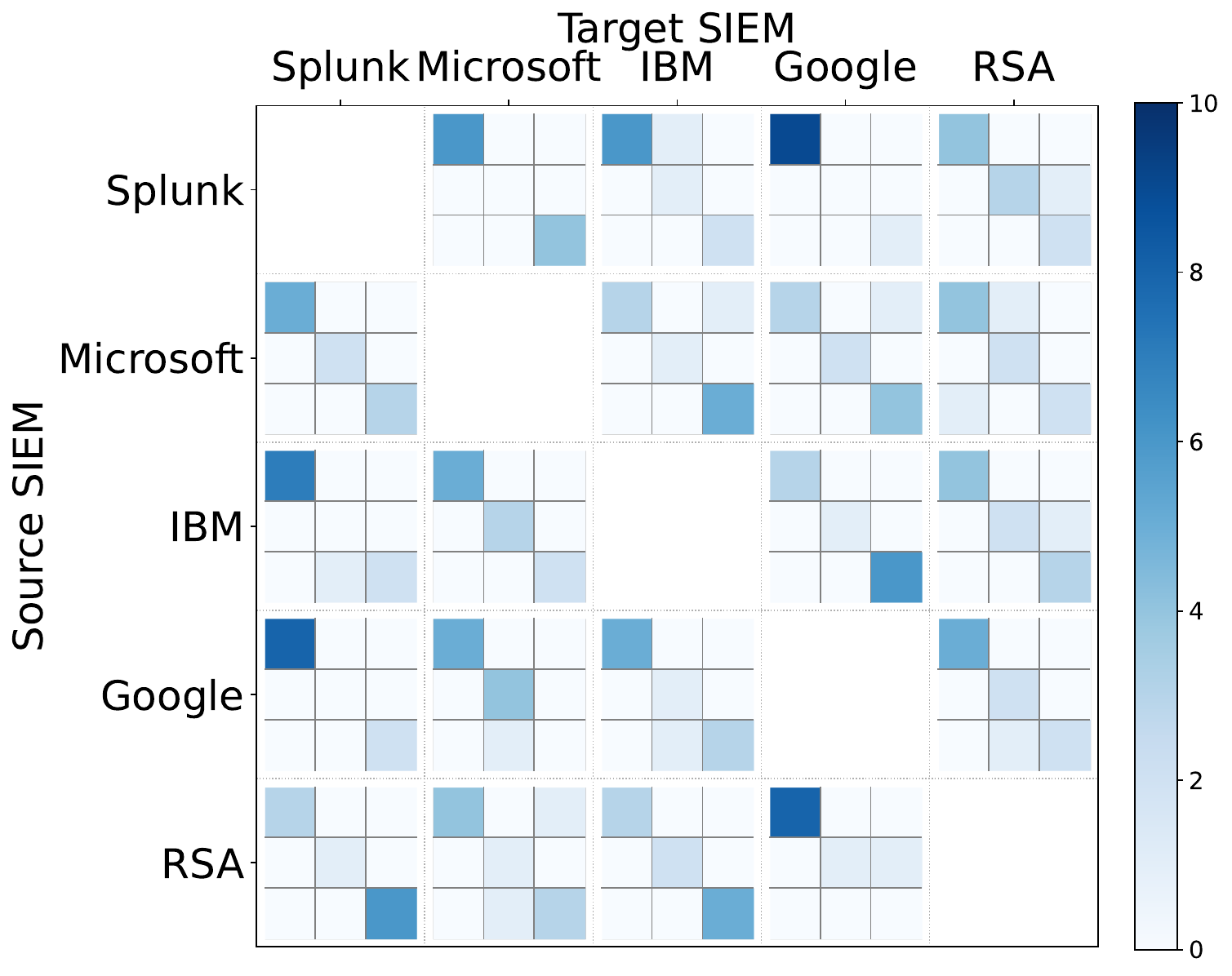}
    \caption{Consistency between LLM-as-a-Judge and human evaluation for heterogeneous SIEM rule conversions. Each off-diagonal cell corresponds to one source–target conversion and contains a 3×3 confusion matrix: rows are human judgments (\name better / Tie / Baseline better), and columns are LLM judgments. Diagonal entries show agreement, while off-diagonal entries show mismatches.}  
    \label{fig:confusion_matrix}
\end{figure}

\noindent We adopt a scoring scheme ranging from 0 to 1 for each evaluation dimension. We use relative scores between outputs under the same prompt in the baseline and \name, instead of absolute scores. 
To mitigate evaluation bias, we adopted a human-aligned iterative evaluation protocol, where an experienced human expert and an LLM jointly refined prompts to ensure consistent evaluation standards. We categorize each pairwise comparison into three possible outcomes: ``ARuleCon better'', ``Tie'', and ``Baseline better''. 
We define inter-rater agreement as a match in relative preference; for example, both the human and the LLM preferring \name over the corresponding vanilla LLMs (baseline) is considered consistent, regardless of exact numerical differences.
Under this definition, the inter-rater agreement results are visualized in Figure~\ref{fig:confusion_matrix}. The agreement reaches a Cohen’s Kappa~\cite{Cohen-Kappa} score larger than 0.885, indicating strong consistency between human judgment and LLM-as-a-judge evaluation.

\begin{figure}[t]
\setlength{\abovecaptionskip}{0pt}
\setlength{\belowcaptionskip}{0pt}
\footnotesize
\centering
\includegraphics[width=0.45\textwidth]{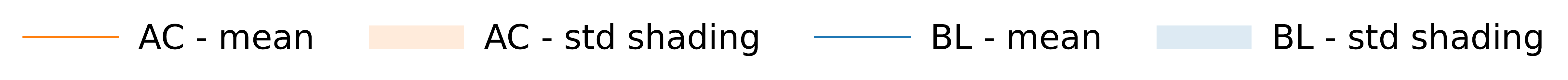}
\vspace{0pt}
\subfigure[\texttt{GPT5}]{\includegraphics[width=0.15\textwidth]{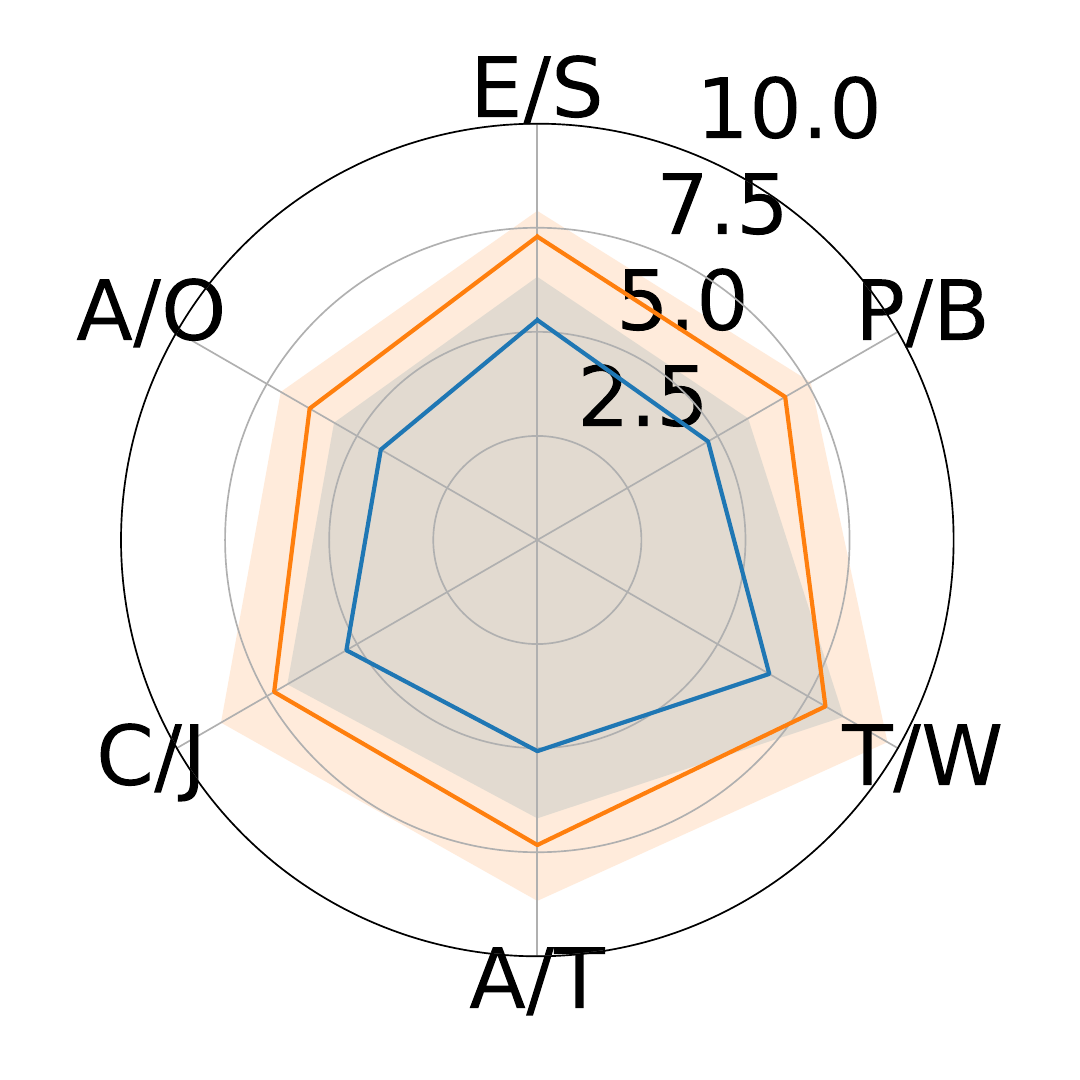}}
\subfigure[\texttt{DeepSeek-V3}]{\includegraphics[width=0.15\textwidth]{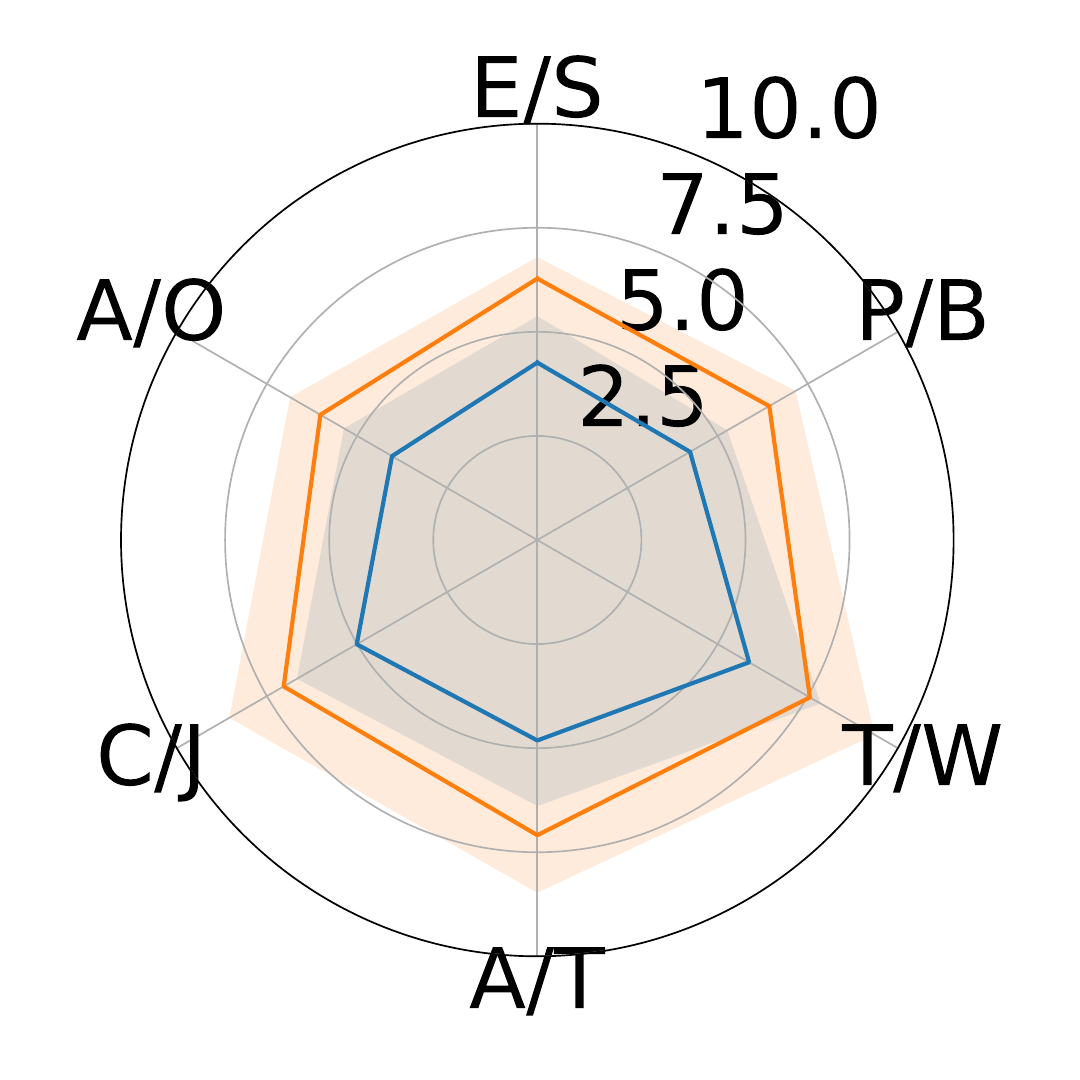}}
\subfigure[\texttt{LLaMa-3}]{\includegraphics[width=0.15\textwidth]{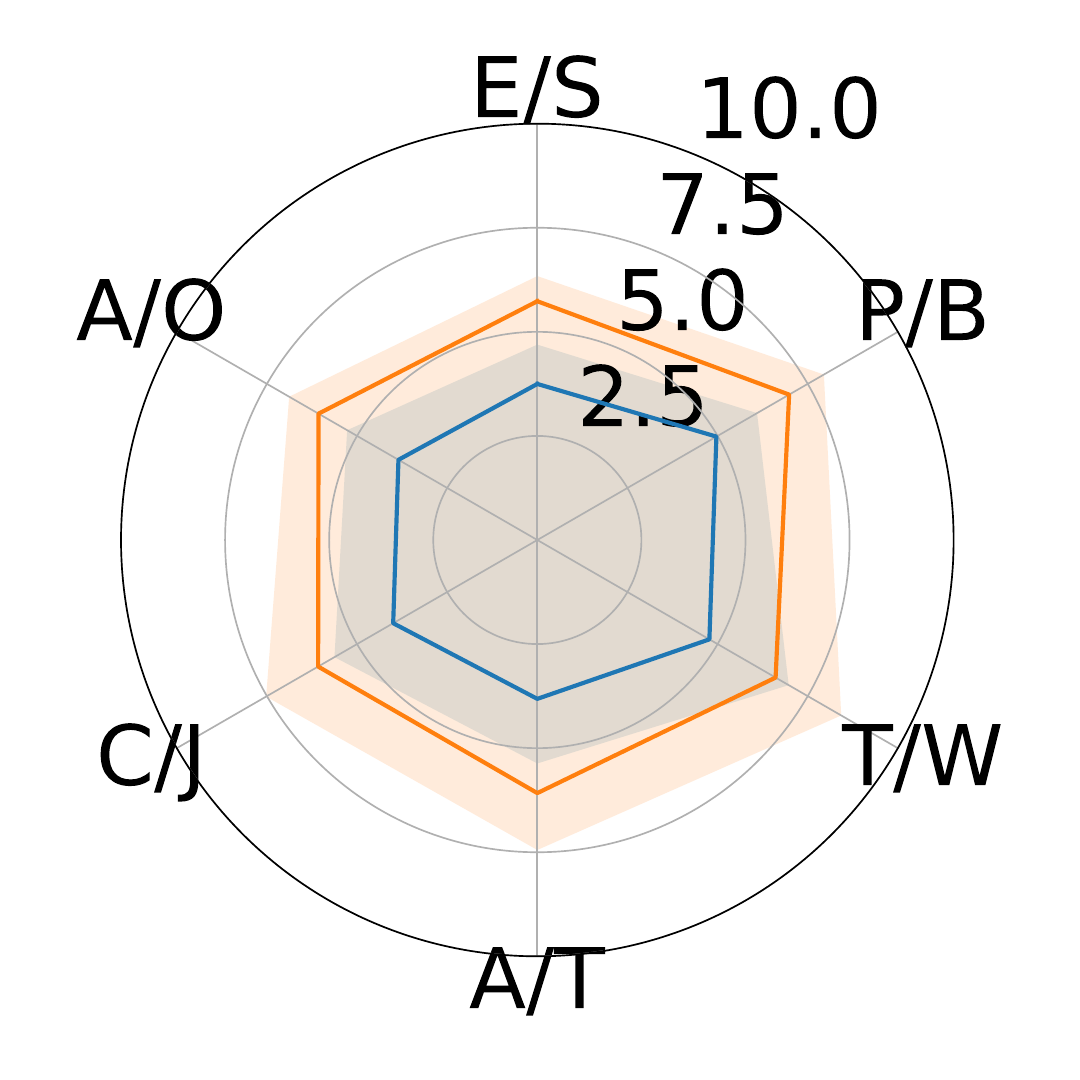}}

\caption{Semantic evaluation of \name.}
\label{fig:radar}
\end{figure}

\noindent\textbf{Results.} The radar chart in Figure~\ref{fig:radar} illustrates the results of the LLM-based evaluator. 
This plots reveal that \name consistently surpasses the baseline across all six semantic facets, with the most pronounced gains in \textit{Predicate \& Boolean Logic} and \textit{Alert Semantics/Outcome}. GPT-5 (Figure~\ref{fig:radar}a) demonstrates the most balanced and stable improvements, DeepSeek (Figure~\ref{fig:radar}c) achieves noticeable gains but with smaller margins in \textit{Scope/Schema} and \textit{Temporal}, while LLaMA (Figure~\ref{fig:radar}e) shows narrower coverage and occasional regressions, especially in \textit{Aggregation \& Thresholding}. 
We show every source-target cases in Figure~\ref{fig:radardetails}, which further confirm these observations: conversions involving Splunk and RSA NetWitness preserves more semantic validity. 
Overall, these results validate that \name achieves consistent semantic fidelity across heterogeneous SIEMs and remains effective regardless of the underlying backbone LLM.

\begin{figure}[htbp]
\setlength{\abovecaptionskip}{0pt}
\setlength{\belowcaptionskip}{0pt}
\centering
\includegraphics[width=0.70\linewidth]{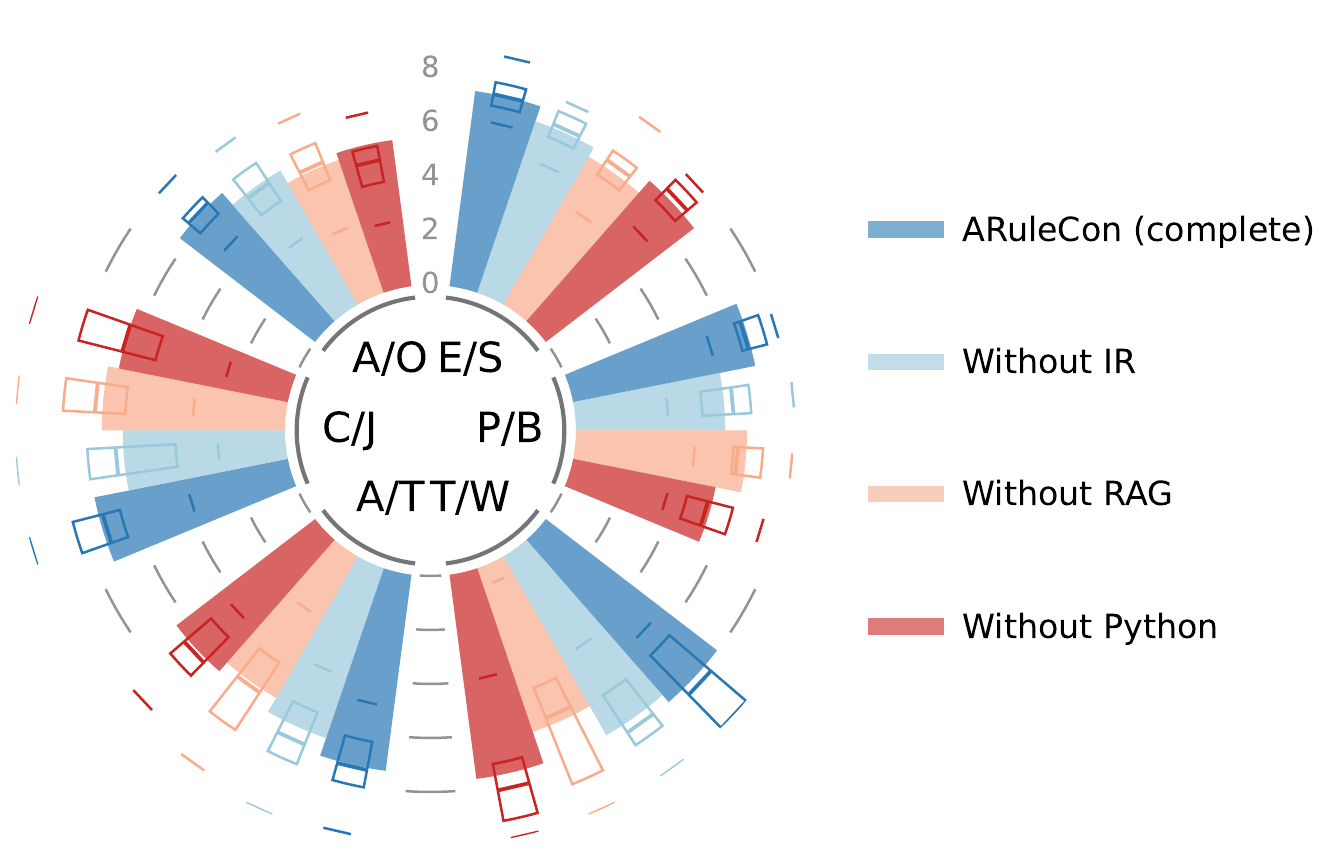}
\caption{Ablation study of semantic metrics on \name.}
\vspace{-2mm}
\label{fig:ablation_polar} 
\end{figure}

\noindent\textbf{Ablation Study.} Figure~\ref{fig:ablation_polar} shows the semantic evaluation from the LLM-as-a-judge. The complete \name consistently outperforms its ablated variants across all six dimensions. Removing IR mainly reduces performance on Aggregation \& Thresholding (A/T) and Correlation \& Joins (C/J), since these require explicit step decomposition. Without Agentic RAG, scores on Predicate \& Boolean Logic (P/B) and Event Scope \& Schema Mapping (E/S) drop more, likely due to weaker field alignment and operator mapping. Excluding Python checks most affects Alert Semantics \& Outcome (A/O), as execution mismatches cannot be detected. Overall, the results indicate that each component contributes complementary strengths, and their combination is crucial for robust semantic consistency.

\subsection{Case Analysis}~\label{app:temporal-windows-thresholds} 

\noindent\textbf{Nested Query Structures.} 
A common difficulty in cross-SIEM rule translation emerges when rules contain nested query logic. 
Splunk allows such patterns through subsearches (\colorbox{mygray}{\texttt{[...]}}), enabling analysts to embed one query as the filter condition of another. 
Consider the following detection task: identify outbound network flows to rare non-US destinations, but only for source IPs that have already triggered multiple failed logins within a recent time window. 
This is to detect the realistic attack pattern when adversaries attempt brute-force authentication before establishing external connections. 
In Splunk SPL, this can be expressed with a nested query:

\begin{lstlisting}
index=netflow action=allowed dest_country!=US
| lookup geoip ip as dest_ip output country as dest_country
| where dest_country!="US"
§[ search index=auth action=failure user!=root earliest=-30m
  | stats count as fail_cnt by src_ip
  | where fail_cnt >= 5
  | fields src_ip ]§
| stats dc(dest_country) as unique_countries, values(dest_country) as dest_list by src_ip
| where unique_countries >= 2
\end{lstlisting} 

The nested form in Splunk is challenging to be directly mapped to another SIEM-sules, which lacks subsearch support. 
Chronicle requires each event type to be explicitly declared and joined, with conditions written as correlations over event variables. 
Through the use of IR, the nested logic can be decomposed into a series of sequential steps that separate the construction of suspicious IP sets from the subsequent flow analysis.
For the case of outbound connections from IPs with repeated failed logins, the IR representation is as follows:

\begin{minted}[fontsize=\small, frame=single, breaklines, bgcolor=lightgray!10]{json}
[
  /* ... */
  {
    "keyword": "failed_login_count",
    "param": "src_ip >= 5 within 30m",
    "description": "Identify source IPs with at least 5 failed logins in the last 30 minutes"
  },
  {
    "keyword": "outbound_flow_filter",
    "param": "dest_country != US",
    "description": "Select outbound connections to non-US destinations for those IPs"
  },
  /* ... */
]
\end{minted}

\noindent Based on this structured breakdown, the corresponding Chronicle YARA-L rule can then be written as:

\begin{lstlisting}
rule brute_force_outbound {
  meta:
    description = "Outbound connections to non-US from IPs with >= 5 failed logins in 30m"
  events:
    §$fail e1 = m.system.auth {
      filter: security_result.action = "BLOCK"}§
    $flow e2 = net.flow {
      filter: network.connection_action = "ALLOW"
              and destination.country != "US"}
  match:
    §$fail.src.ip = $flow.src.ip over 30m§
  condition:
    §count($fail) by $fail.src.ip >= 5§
    and count_distinct($flow.destination.country) >= 1
  outcome:
    $fail.src.ip,
    $flow.destination.country
}

\end{lstlisting} 

\noindent\textbf{Temporal Windows \& Thresholds.}
Temporal windows and thresholds are widely used in SIEM rules, for example, to detect repeated failed login attempts within a fixed time range. 
Different SIEMs implement the semantics with different syntactic constructs.     
Splunk discretizes time with \colorbox{mygray}{\texttt{bucket span=…}} and then aggregates via \colorbox{mygray}{\texttt{stats}} (e.g., \texttt{| bucket \_time span=10m | stats count by …, \_time}). Microsoft Sentinel (KQL) instead bins timestamps with \colorbox{mygray}{\texttt{bin(TimeGenerated, …)}} inside summarize (e.g., \texttt{... | summarize count() by ..., bin(TimeGenerated, 10m)}). Beyond keyword differences (\colorbox{mygray}{\texttt{bucket}} vs. \colorbox{mygray}{\texttt{bin}}), they also diverge on time column names (\colorbox{mygray}{\texttt{\_time}} vs. \colorbox{mygray}{\texttt{TimeGenerated}}) and the placement of filters relative to aggregation. 

Consider the following Splunk SPL rule, which detects potential DNS tunneling behavior. The rule flags any host that, within a 10-minute window, issues at least 100 DNS to rare top-level domains (such as .xyz, .pw), across three or more distinct second-level domains. 
\begin{lstlisting}
index=dns sourcetype=dns:query
| eval tld=lower(replace(query, ".*\\.([A-Za-z0-9-]+)$", "\1"))
| where tld IN ("xyz","pw","top")
| eval sld=lower(replace(query, ".*\\.([A-Za-z0-9-]+\\.[A-Za-z0-9-]+)$", "\1"))
| §bucket _time span=10m§
| §stats count as qps, dc(sld) as sld_cnt, values(sld) as sld_set by host, _time§
| where qps >= 100 AND sld_cnt >= 3
\end{lstlisting}  

Through Agentic RAG, the system simulates how an analyst would interact with vendor documentation during translation. From the IR intent, the agent first generates exploratory keywords such as \textit{“KQL time-window aggregation”} and \textit{“KQL distinct count”}. The retrieval step then surfaces the official summarize operator page, which prescribes the use of \texttt{summarize … by bin(TimeGenerated, 10m)}, and the list of supported aggregation functions including \colorbox{mygray}{\texttt{dcount()}} and \colorbox{mygray}{\texttt{make\_set()}}. From these references, the agent acquires concrete knowledge: Splunk’s \texttt{bucket \_time span=30m} should be rewritten as \texttt{bin(TimeGenerated, 30m)}, \colorbox{mygray}{\texttt{dc()}} maps to \colorbox{mygray}{\texttt{dcount()}}. 
The faithful KQL rule produced through this process is shown below.   

\begin{lstlisting}
DnsEvents
| extend TLD = tolower(extract(@".*\.([A-Za-z0-9-]+)$", 1, QueryName))
| where TLD in ("xyz","pw","top")
| extend SLD = tolower(extract(@".*\.([A-Za-z0-9-]+\.[A-Za-z0-9-]+)$", 1, QueryName))
| §summarize qps = count(),
            sld_cnt = dcount(SLD),
            sld_set = make_set(SLD)
    by Computer, bin(TimeGenerated, 10m)§
| where qps >= 100 and sld_cnt >= 3
\end{lstlisting}

\end{document}